\documentclass[fleqn,usenatbib]{rasti}

\usepackage{newtxtext,newtxmath}

\usepackage[T1]{fontenc}

\DeclareRobustCommand{\VAN}[3]{#2}
\let\VANthebibliography\thebibliography
\def\thebibliography{\DeclareRobustCommand{\VAN}[3]{##3}\VANthebibliography}

\usepackage{graphicx}	
\usepackage{amsmath}	
\usepackage{booktabs}
\usepackage{multirow}
\usepackage{makecell}
\usepackage{subcaption}
\usepackage{fix-cm}

\newcommand{\rev}[1]{#1}

\title[Atmospheric Characterisation with Twinkle]{Atmospheric Characterisation with the Twinkle Space Telescope Following Advances from JWST Observations}

\author[Zhang et al.]{Tailong Zhang,$^{1}$\thanks{E-mail: tailong@bssl.space}
Benjamin Wilcock,$^{1}$
Sushuang Ma,$^{2}$
Giovanna Tinetti,$^{1,2}$
Lawrence Bradley,$^{1}$
\newauthor
Ian Stotesbury,$^{1}$
Marcell Tessenyi,$^{1}$
and Jonathan Tennyson,$^{1,3}$
\\
\\
$^{1}$Blue Skies Space Ltd., 69 Wilson Street, London, EC2A 2BB, UK\\
$^{2}$Department of Physics, King’s College London, Strand, London WC2R 2LS, UK \\
$^{3}$Department of Physics and Astronomy, University College London, Gower Street, London WC1E 6BT, UK}

\date{Accepted XXX. Received YYY; in original form ZZZ}

\pubyear{\the\year{}}

\begin{document}
\label{firstpage}
\pagerange{\pageref{firstpage}--\pageref{lastpage}}
\maketitle

\begin{abstract}

The Twinkle Space Telescope is a satellite designed for spectroscopic observations of a wide range of extrasolar and solar system objects. Equipped with a 0.45 m diameter telescope and a spectrometer covering  from 0.5 to 4.5~$\mu$m simultaneously, Twinkle will be launched in a sun-synchronous, low-Earth orbit, and it is expected to operate for seven years. Twinkle is developed, managed and operated by Blue Skies Space (BSSL), a space science data company whose vision is to accelerate and expand the availability of new, high-quality datasets to researchers worldwide, complementing the space-observatories delivered by government space agencies. Over its life-time, Twinkle will conduct large-scale survey programs. The scientific objectives and observational strategy of these surveys are defined by researchers who join the Science Team. Leveraging  advances made possible by recent observations with the James Webb Space Telescope, we present here updated simulations evaluating Twinkle’s observational capabilities in the context of exoplanet atmospheres. Through retrieval analyses of HD\,209458\,b, WASP-107\,b, GJ\,3470\,b, and 55\,Cnc\,e, we demonstrate how increasing observational investment enhances the retrieval of atmospheric parameters and molecular abundances. Our sensitivity study highlights Twinkle’s capability to detect less abundant/detectable molecules depending on the observing strategies adopted. This work provides practical guidance for developing targeted observational strategies to maximise Twinkle’s scientific return.

\end{abstract}




\section{Introduction}
\label{sec:intro}
The field of exoplanetary science has experienced unprecedented growth, with the number of confirmed exoplanets now exceeding 5,800. A significant proportion of these detections -- approximately 4,000 -- has been made using transit photometry from space-based surveys such as Kepler, K2, and TESS \citep{2010Sci...327..977B, 2014PASP..126..398H, 2014SPIE.9143E..20R}. Since the retirement of the Kepler/K2 mission in 2018, the TESS mission, operational since 2018, has significantly advanced planet discovery efforts. As of 12 March 2025, TESS has identified 7,525 TESS Objects of Interest (TOIs). After excluding false positives, 5,253 candidates remain confirmed by the TESS Follow-up Observing Program Working Group (TFOPWG) \citep{2019AAS...23314009A}. This wealth of discoveries of transiting planets around bright stellar hosts has  been a game changer for atmospheric characterisation studies using transit, eclipse and phase-curve spectroscopy.

The European Space Agency's PLATO, due for launch in 2026,  will aim at detecting small transiting planets (above $2 R_{Earth}$)  around bright stars ($\leq$11 mag), including terrestrial planets in the habitable zone of solar-like stars \citep{2025ExA....59...26R}.  Similarly, the Earth 2.0 (ET) mission in transit mode will continuously monitor over 2 million FGKM dwarfs in the original Kepler field and its neighbouring fields for four years to search for new planets, including terrestrial-like planets, across a wide range of orbital periods \citep{2022arXiv220606693G, 2024SPIE13092E..18G}.

Since the first atomic, ionic and molecular species were revealed in exoplanetary atmospheres \citep[e.g.,][]{2002ApJ...568..377C, 2003Natur.422..143V, 2004ApJ...604L..69V, 2008ApJ...673L..87R, 2007Natur.448..169T, 2008Natur.452..329S, 2009ApJ...704.1616S, 2008Natur.456..767G}, the interest in studying the chemistry and structure of exoplanet atmospheres has progressively become more widespread. As a result,  tens of exoplanetary atmospheres have been observed with the Spitzer and Hubble Space Telescopes and with ground-based facilities in the past couple of decades.  These measurements have guided our initial understanding of the atmospheric composition, dynamics and thermal and scattering properties for a variety of atmospheres in very diverse environments  \citep[e.g.,][]{2007Natur.447..183K, 2006Sci...314..623H, 2012ApJ...747L..20M, 2014Sci...346..838S, 2016Natur.532..207D}. 

In particular, the WFC3 G141 instrument, covering 1.1 to 1.7 microns, has been central to the hunt for key  molecular species such as H\textsubscript{2}O, CH\textsubscript{4},  TiO, VO and clouds for a long list of very diverse exoplanets \citep[e.g.,][]{2011ApJS..193...27W, 2012ApJ...747...35B, 2013ApJ...774...95D, 2014Natur.513..526F, 2016ApJ...820...99T, 2018ApJ...855L..30A}. 

While this effort has been pursued through a number of complementary techniques, including  direct imaging and high-resolution spectroscopy from the ground \citep{2010Natur.468..669B,
2010Natur.465.1049S, 2018Sci...362.1388N, 2015Sci...350...64M, 2017A&A...605L...9C, 2019A&A...623L..11G}, transit, eclipse,  phase-curve spectroscopy and multi-band photometry from space observatories have enabled large-scale atmospheric studies  throughout the years \citep{2016Natur.529...59S, 2018AJ....155..156T, 2023ApJS..269...31E, 2022ApJS..260....3C,2025ApJS..276...70S, 2025AJ....169...32D}.

The recent launch of the James Webb Space Telescope (JWST) has offered unprecedented insights into exoplanet atmospheres, spanning wavelengths from 0.5 to 14~$\mu$m \citep[e.g.,][]{2023Natur.614..649J, 2023ApJ...956L..13M, 2024ApJ...963L...5X, 2024NatAs...8.1008C, 2024Natur.630..609H, 2024AAS...24334304B}. JWST has refined measurements of molecular abundances, thermal profiles, atmospheric dynamics, and cloud/haze characteristics. Despite these significant advances, JWST's broad scientific remit -- including galactic and extragalactic astrophysics  and solar system studies -- limits its dedicated observation time for exoplanetary research.

Addressing this limitation, dedicated missions such as ESA's Ariel (scheduled for launch in 2029) have been developed. Ariel aims to conduct a comprehensive chemical survey of exoplanetary atmospheres using transit spectroscopy over the spectral range of 0.5–7.8~$\mu$m, thereby deepening our understanding of how atmospheric chemistry correlates with planetary characteristics \citep{2018ExA....46..135T, 2022EPSC...16.1114T}.


The Twinkle Space Telescope,  managed and operated by Blue Skies Space Ltd. (BSSL), will be dedicated to provide spectroscopic data for various science themes developed by its members \citep{jason2016twinkle, savini2018twinkle, edwards2019remote, edwards2019small, archer2020sustainable, 2022SPIE12180E..33S, 2024SPIE13092E..13S}. Equipped with a 0.45-metre telescope and a spectrometer covering simultaneously from 0.5 to 4.5~$\mu$m, Twinkle is capable of providing spectra for a wide range of targets such as brown dwarfs, stars, protoplanetary disks, Solar System objects and exoplanets. Two large survey programmes  focusing on extrasolar targets and Solar System objects are being planned during the first three years of Twinkle's expected seven years operations   \citep{2022SPIE12180E..33S}. While BSSL is responsible for the satellite, the data analysis tools and facilitating the survey planning process, researchers who joins the survey program (Science Team) will decide the scientific objectives and observational strategy. Twinkle Science Team members have been tasked with delivering a comprehensive and scientifically optimised observation plan and list of targets. 

Based on the existing scientific themes proposed by the Twinkle Science Team members, a large fraction of the survey time will be allocated to extrasolar objects and the atmospheric characterisation of diverse types of exoplanets is expected to be a significant part of the survey. This paper presents new simulations to evaluate Twinkle’s potential for atmospheric characterisation in the context of more recent findings by JWST and discusses possible approaches  to optimise the observational strategy.

Many of these upcoming missions have spectral ranges that significantly overlap with JWST, making JWST's observations an invaluable reference for simulating their expected performance. The combination of JWST’s comprehensive spectral coverage and its recent discoveries offers essential benchmarks that drive our updated simulations of Twinkle’s observational capabilities. In this study, we incorporate the latest JWST-calibrated atmospheric models into Twinkle’s simulation framework, leading to more accurate predictions of signal-to-noise ratios (SNR) and a refined assessment of the detectability of key atmospheric features.

Previous studies by \citet{2019ExA....47...29E} evaluated  Twinkle's expected performance to study the atmospheric composition and structure of three iconic exoplanets, HD\,209458\,b, GJ\,3470\,b, and 55\,Cnc\,e. However, due to lack of IR spectroscopic  data  pre-JWST, those analyses were based on theoretical models and simplified assumptions. Recent JWST observations have revealed complexities in atmospheric chemistry,  thermal and cloud structures  that necessitate an update to these models \citep[e.g.,][]{2024ApJ...963L...5X, 2024ApJ...970L..10B, 2024Natur.630..609H}. 
More specifically,  recent JWST observations of HD\,209458\,b by \citet{2024ApJ...963L...5X} have provided a high-quality transmission spectrum from 2.3 to 5.1~$\mu$m, revealing pronounced absorption features from CO\textsubscript{2} and H\textsubscript{2}O. Based on thermochemical equilibrium assumptions, these results suggest a supersolar metallicity and a  low C/O ratio ($\sim$0.11).  
GJ\,3470\,b has emerged as a key target of interest  due to its relatively transparent atmosphere and the detection of SO\textsubscript{2} in its atmosphere, reinforcing the evidence of disequilibrium chemistry in lower-mass  exoplanet atmospheres \citep{2024ApJ...970L..10B}.  JWST's eclipse observations of the super-Earth 55\,Cnc\,e by \citet{2024Natur.630..609H} have confirmed the presence of a volatile-rich atmosphere, potentially maintained by a magma ocean. In addition to these three exoplanets, we have 
added here the study of WASP-107\,b:
\citet{2024Natur.630..836W} have identified the presence of several molecular species in the atmosphere of this planet, including H\textsubscript{2}O, CH\textsubscript{4}, CO, CO\textsubscript{2}, SO\textsubscript{2}, and NH\textsubscript{3} when combining HST and JWST data.

\begin{table*}
\centering
\caption{Exoplanets and their properties considered in this work. HD\,209458\,b \citep{2017AJ....153..136S, 2024ApJ...963L...5X}; WASP-107\,b \citep{2017A&A...604A.110A, 2021yCat..51610070P}; GJ\,3470\,b \citep{2016MNRAS.463.2574A}; 55\,Cancri\,e \citep{2024Natur.630..609H}.}
\label{tab:exoplanets}
\begin{tabular}{lcccc}
\hline
 & HD\,209458\,b & WASP-107\,b & GJ\,3470\,b & 55\,Cancri\,e \\
\hline
Planet Type & Hot Jupiter & Warm Neptune & Warm Neptune & Super Earth \\
Planet Mass ($M_{\mathrm{J}}$) & $0.73 \pm 0.04$ & $0.096 \pm 0.005$ & $0.035 \pm 0.005$ & $0.0251 \pm 0.001$ \\
Planet Radius ($R_{\mathrm{J}}$) & $1.39 \pm 0.02$ & $0.94 \pm 0.02$ & $0.37 \pm 0.016$ & $0.1673 \pm 0.0026$ \\
T\textsubscript{eq} (K) & $1140^{+71}_{-83}$ & $770 \pm 60$ & $600 \pm 96.5$ & $1958 \pm 15$ \\
Stellar Magnitude (V/K) & 
\makecell{7.65 $\pm$ 0.03 (V)\\6.308 $\pm$ 0.026 (K)} &
\makecell{11.592 $\pm$ 0.208 (V)\\8.637 $\pm$ 0.023 (K)} &
\makecell{12.332 $\pm$ 0.016 (V)\\7.989 $\pm$ 0.023 (K)} &
\makecell{5.95 $\pm$ 0.023 (V)\\4.015 $\pm$ 0.036 (K)} \\
JWST Availability ($\mu$m) & 2.3--4.5 & 2.4--4.5 & 2.45--4.97 & 3.94--12 \\
HST Wavelength Range ($\mu$m) & 0.5--0.9; 1.1--1.7 & 0.8--1.6 & \makecell{0.524--1.027;\\1.1--1.7} & -- \\
\hline
\end{tabular}
\end{table*}

This paper first reviews Twinkle's instrumental capability, the models and methodology used in estimating Twinkle's science performance in Section \ref{sec:methods}. It updates the radiometric model employed to simulate the noise characteristics (Section \ref{sec:radiometric_snr}) and the list of potential target candidates within Twinkle's field of view and details. Section \ref{sec:exoplanets_retrieval} describes the atmospheric forward and inverse models obtained with the open-source exoplanet atmospheric retrieval framework \texttt{TauREx\,3}.
Simulations and retrieval results are presented in Section \ref{sec:results}, highlighting Twinkle's 
ability to detect molecular species and other atmospheric properties across a diverse exoplanetary sample and observational strategies, based on recent JWST advances. Finally, Section \ref{sec:discussion} discusses the results, addresses study limitations, and proposes directions for future simulation work.

\section{Methods}
\label{sec:methods}
\subsection{Twinkle's Radiometric Model}
\label{sec:radiometric_snr}

Twinkle will observe spectra simultaneously across two channels covering the wavelength range from 0.5 to 4.5~$\mu$m. Channel 0 operates from 0.5 to 2.43~$\mu$m with a resolving power ($R = \frac{\lambda}{\Delta\lambda}$) of up to 70, and channel 1 operates from 2.43 to 4.5~$\mu$m with a resolving power of up to 50. 

Twinkle's radiometric performance is assessed using the Twinkle Radiometric Tool \citep{2022SPIE12180E..33S}, adapted from the open-source radiometric software ExoRad2 \citep{2023JOSS....8.5348M}. The radiometric model simulates observational and instrumental noise contributions, including photon noise, instrument emission, detector dark current, read noise, and zodiacal background. This model has been refined to incorporate recent advances in instrument design \rev{(as elaborated in \citet{2024SPIE13092E..13S}). Detailed noise contributions for each target simulated in this study are presented in Appendix~\ref{appendix:noise}. It is worth noting that the scope of this study covers only radiometric simulation, without detailed time-domain simulation, which will be addressed in a forthcoming paper. However, an average jitter value derived from ground tests is applied.}
\rev{The assumed instrument parameters are summarised in
Table~\ref{tab:instrument_params} of Appendix~\ref{appendix:noise}.}

The radiometric tool calculates realistic noise estimates, which in turn allows determination of the number of observations necessary to reach a specific SNR through observation stacking. 

The cumulative SNR achieved by combining multiple observations can be expressed as:

\begin{equation}
    SNR_N = \sqrt{N} \times SNR_1
    \label{eq:number_transits}
\end{equation}
where $N$ is the number of observations \rev{(each corresponding to a single transit or eclipse event)}, and $\text{SNR}_1$ is the signal-to-noise ratio (SNR) of a single observation. The signal is derived from the amplitude ($A_\mathrm{p}$) of spectral features in transit spectra, approximated by:
\begin{equation}
    A_\mathrm{p} = \frac{2R_\mathrm{p} \times z}{R_\mathrm{s}^2}
\end{equation}
Here, $R_\mathrm{p}$ and $R_\mathrm{s}$ represent the planetary and stellar radii, respectively, and $z$ denotes atmospheric thickness. The atmospheric thickness is typically estimated as $nH$, with $n$ ranging from 3 to 5 for hydrogen/helium (H$_2$/He)-dominated atmospheres. The scale height $H$ is defined by:
\begin{equation}
    H = \frac{k_\mathrm{B}T_\mathrm{eq}}{\mu g}
\end{equation}
where $k_\mathrm{B}$ is Boltzmann’s constant, $T_\mathrm{eq}$ the equilibrium temperature of the planet, $g$ the planet’s surface gravity, and $\mu$ the mean molecular weight of the atmosphere (usually assumed as 2.3 for H$_2$/He atmospheres but refined using recent JWST data in this study).

 Previously, due to limited observational constraints, atmospheric thickness ($z$) was set as 1, 3, or 5 scale heights, with 5 scale heights being a commonly used assumption \citep[e.g.,][]{2019ExA....47...29E, 2024MNRAS.530.2166B}. However, with the aid of recent data release from JWST and its overlapping wavelength coverage with Twinkle, constraints on atmospheric composition have improved, enabling more precise estimates of atmospheric scale height and thickness above the optically thick continuum. Thus, we can derive a more realistic SNR and resolution needed for Twinkle to better retrieve the atmospheric chemical composition, and consequently, determine the number of observations required to achieve a given SNR. It is worth noting that the radiometric model employed in this study calculates Twinkle's observational errors assuming complete transit/eclipse coverage. However, observational availability of targets are influenced by Earth obstruction due to Twinkle's orbital configuration, causing Twinkle to sometimes capture partial transits or eclipse events, similar to the situation encountered by CHEOPS~\citep{2021ExA....51..109B}. \rev{The extent of transit or eclipse coverage depends on the target's declination, the satellite's orbital phase, and the ratio of the planet's orbital period to Twinkle's orbital period. Partial transits can still yield scientifically useful data but at reduced observing efficiency compared with full-coverage events. A detailed quantification of this efficiency factor, incorporating the effects of Earth obstruction and target-specific orbital geometry, will be addressed in a forthcoming time-domain simulation study.}

While for some exoplanets, the number of required observations is high, it is important to note that lowering spectral resolution can increase the SNR per wavelength bin and thus \rev{achieve the desired SNR with fewer stacked observations}. Although \rev{post-observation rebinning} is widely used \citep[e.g.,][]{2014ApJ...793L..27K, 2014Sci...346..838S}, artificially reducing resolution in simulations introduces \rev{a loss of spectral information} along the wavelength axis, particularly around spectral peaks. To avoid this, this study retrieves spectra at Twinkle's \rev{full native} resolution. \rev{After launch, rebinning applied to real observational data---rather than to simulated spectra---will mitigate this limitation, as it preserves the full information content of the original observations.}


\subsection{Twinkle's Target Candidates}
\label{sec:target}
Twinkle will be placed in a 1200 km, sun-synchronous low-Earth orbit, with a 6am Local Time of Ascending Node (LTAN). The spacecraft's observational capabilities are restricted by the Field of Regard (FoR) (a 40 degree cone around the anti-Sun vector), solar \& lunar exclusion constraints and Earth obstruction across different seasons \citep{2022SPIE12180E..33S}.

To systematically assess target observability, we employed the BSSL Twinkle Radiometric Tool to simulate observational noise for all confirmed exoplanets within Twinkle’s FoR possessing well-defined stellar and planetary parameters. 
Planetary and stellar data from the NASA Exoplanet Archive are taken as input into the radiometric tool to simulate noise levels for single transits or eclipses at Twinkle’s maximum resolving power.
For TESS Objects of Interest (TOIs) lacking direct planetary mass measurements and thus not able to be processed with Radiometric Tool, masses were estimated using the \rev{\textit{Forecaster}} tool \citep{2017ApJ...834...17C}. Approximately 200 TOIs within Twinkle’s FoR were subsequently excluded due to unavailable stellar mass data. Additionally, \rev{planets with radii between 8 and 22 Earth radii fall outside the calibrated range of \textit{Forecaster}, and the} initial mass-radius relationships for gas giants proved inconsistent with current empirical data. Hence, planetary masses for these TOIs were recalculated based on a log-normal distribution fitted to known transiting exoplanet populations, as detailed in Figure~\ref{fig:mass_prediction} and further described in Appendix~\ref{sec:A_mass}. \rev{TOIs lacking sufficient stellar or planetary data for reliable mass estimation were excluded from the candidate list.}

Our primary focus was on exoplanets with H$_2$/He dominated atmospheres; thus, we assumed a mean molecular weight of 2.3 for initial assessments. Atmospheric thicknesses equivalent to 3 atmospheric scale heights were adopted uniformly for all candidate targets. Exoplanets predicted to achieve median SNR levels of 3, 5, or 7 within 30 transits or eclipses at Twinkle's native spectral resolution were considered suitable candidates for detailed atmospheric characterisation. \rev{The threshold of 30 observations represents an assumed upper limit per target, determined by Twinkle's total available observing time, the number of targets in the survey, and the accessibility constraints imposed by the satellite's orbit over the mission lifetime. This choice of upper limit is an operational planning decision rather than a constraint imposed by Twinkle's instrumental capabilities.}

\subsection{Simulated Exoplanets and Spectral Retrievals}
\label{sec:exoplanets_retrieval}

The detection of molecular species and characterisation of chemical compositions and abundances in exoplanet atmospheres remain central objectives of transit spectroscopy. Achieving these goals offers critical insights into planetary processes, formation histories, and potential habitability. Twinkle's extensive and continuous spectral coverage from optical to near-infrared wavelengths includes absorption features for many previously detected and theoretically predicted molecules. In this study, we conduct atmospheric retrieval simulations on selected exoplanets (listed in Table \ref{tab:exoplanets}), chosen from Twinkle’s candidate list based on the availability of JWST data. These planets represent a diverse range of types, aligning with Twinkle’s broader exoplanetary science themes, and facilitating a comprehensive evaluation of Twinkle’s capabilities informed by recent JWST findings.

We employed the open-source retrieval code \texttt{TauREx 3.2.2} \citep{al2021taurex} to simulate forward models for transmission and emission spectra and to perform spectral retrieval analyses. Stellar spectra for each target star were simulated using the PHOENIX stellar library \citep{2013A&A...553A...6H}. Chemical profiles were generated using the FastChem equilibrium chemistry plugin \rev{\citep{2018MNRAS.479..865S}} for equilibrium assumptions, supplemented by TauREx's \rev{free chemistry retrieval mode. In this mode, the volume mixing ratio of each molecule is treated as a free parameter, constant with altitude, and retrieved independently without imposing chemical equilibrium constraints \citep{al2021taurex}. While this approach does not explicitly model disequilibrium chemistry, it provides a model-agnostic framework that can capture departures from equilibrium, such as those arising from vertical quenching or photochemistry}. All atmospheric models assumed a plane-parallel geometry comprising 100 layers, including molecular absorption, Rayleigh scattering, grey clouds, and collision-induced absorption (CIA). For all exoplanets, CIA processes involving H\textsubscript{2}-H\textsubscript{2} and H\textsubscript{2}-He were considered, with additional contributions from N\textsubscript{2}-N\textsubscript{2} and O\textsubscript{2}-O\textsubscript{2} specifically included for 55\,Cnc\,e. Molecular opacity cross-sections were sourced from the ExoMol database \citep{2024JQSRT.32609083T}, while CIA data were taken from the HITRAN database \citep{2019Icar..328..160K}. 

 Retrieval analyses were performed using the Bayesian MultiNest algorithm \citep{2009MNRAS.398.1601F}, employing 1000 live points and an evidence tolerance threshold of 0.5. \rev{Table~\ref{tab:priors} summarises the prior distributions adopted for all retrieval parameters.} Non-informative priors for molecular abundances were uniformly set on a log scale within the range of $10^{-12}$ to $10^{-1}$, while other planetary parameters utilised priors constrained by previous literature of specific planets.

\begin{table}
\renewcommand{\arraystretch}{1.3}
\centering
\caption{\rev{Prior distributions adopted in the free-chemistry retrievals. All molecular log-abundances share a common LogUniform prior; planet-specific parameters are listed separately. $R_p$ is expressed as a multiplicative factor of the nominal planetary radius.}}
\label{tab:priors}
\begin{tabular}{lll}
\hline
\textbf{Parameter} & \textbf{Prior type} & \textbf{Range} \\
\hline
\multicolumn{3}{c}{\textit{Common to all planets}} \\
\hline
All molecular log-VMRs & LogUniform & $[10^{-12},\;10^{-1}]$ \\
\hline
\multicolumn{3}{c}{\textit{HD\,209458\,b}} \\
\hline
$T$ (K) & Uniform & $[100,\;3000]$ \\
$R_p$ factor & Uniform & $[0.5,\;2.0]$ \\
$\log P_{\mathrm{cloud}}$ (Pa) & LogUniform & $[10^{-2},\;10^{7}]$ \\
\hline
\multicolumn{3}{c}{\textit{WASP-107\,b}} \\
\hline
$T$ (K) & Uniform & $[100,\;2000]$ \\
$R_p$ factor & Uniform & $[0.2,\;2.0]$ \\
$\log P_{\mathrm{cloud}}$ (Pa) & LogUniform & $[10^{-2},\;10^{5}]$ \\
\hline
\multicolumn{3}{c}{\textit{GJ\,3470\,b}} \\
\hline
$T$ (K) & Uniform & $[10,\;1500]$ \\
$R_p$ factor & Uniform & $[0.2,\;2.0]$ \\
$\log P_{\mathrm{cloud}}$ (Pa) & LogUniform & $[10^{0},\;10^{6}]$ \\
\hline
\multicolumn{3}{c}{\textit{55\,Cnc\,e}} \\
\hline
$T_{\mathrm{surface}}$ (K) & Uniform & $[1000,\;4000]$ \\
$T_{\mathrm{top}}$ (K) & Uniform & $[1000,\;2000]$ \\
$R_p$ factor & Uniform & $[0.01,\;0.5]$ \\
$\log P_{\mathrm{surface}}$ (Pa) & LogUniform & $[10^{3},\;10^{8}]$ \\
$\log P_{\mathrm{top}}$ (Pa) & LogUniform & $[10^{0},\;10^{5}]$ \\
\hline
\end{tabular}
\end{table}

\rev{To quantify the detectability of individual molecular species, we employ the Likelihood Ratio Test (LRT) following the methodology described by \citet{2013MNRAS.436.2974T}. To guard against unreliable MAP estimates for poorly constrained parameters, we perform the LRT using both the MAP and posterior median values; a molecule is classified as a ``Detection'' only if both tests independently exceed the $3\sigma$ threshold. The full LRT procedure is detailed in Appendix~\ref{appendix:lrt}. The more conservative (lower) detection significance is reported in the results tables; molecules below the $3\sigma$ threshold are classified as ``Upper Limit''.}

To incorporate recent JWST observational findings, chemical compositions and planetary parameters for each simulated exoplanet were based on best-fit solutions reported in recent literature, alongside equilibrium chemistry predictions. Forward modelling and retrieval analyses were performed based on simulated transmission and emission spectra incorporating predicted Twinkle observational errors.

For HD\,209458\,b, recent JWST observations by \citet{2024ApJ...963L...5X} were combined with HST data from \citet{2025ApJS..276...70S}. \rev{The simulated spectrum was aligned with the observed data by applying a vertical offset to reconcile differences in absolute transit depth calibration between the HST and JWST instruments at overlapping wavelengths. Molecular abundances and temperature profile parameters were then fine-tuned to optimise the fit to the combined HST+JWST dataset.} Similarly, to assess Twinkle’s capabilities for WASP-107\,b, a simulated spectrum covering Twinkle’s wavelength coverage (0.5–4.5~$\mu$m) was constructed using HST and JWST NIRSpec data as benchmarks. The atmospheric model included known molecular constituents identified by \citet{2024Natur.630..831S}, specifically H\textsubscript{2}O, CH\textsubscript{4}, NH\textsubscript{3}, CO, CO\textsubscript{2}, SO\textsubscript{2}, and H\textsubscript{2}S, alongside potential SiO detection proposed by \citet{2025arXiv250407823M}, and optical absorbers Na and K, with uniform abundances assumed.

For GJ\,3470\,b, we simulated Twinkle’s performance by closely aligning spectral data with JWST and HST observations across the 0.5–4.5~$\mu$m range. The atmospheric composition model incorporated nine primary molecules (H\textsubscript{2}O, CH\textsubscript{4}, CO\textsubscript{2}, CO, SiO, SO\textsubscript{2}, NH\textsubscript{3}, HCN, H\textsubscript{2}S), with atmospheric temperature fixed at 600 K and a deep cloud deck placed at 3000 Pa to highlight molecular absorption signatures. For the emission spectra of 55\,Cnc\,e, we employed atmospheric assumptions consistent with \citet{2024Natur.630..609H}, featuring predominant CO\textsubscript{2} and CO composition alongside trace species (C\textsubscript{2}H\textsubscript{2}, HCN, O\textsubscript{2}, and N\textsubscript{2}) with constant mixing ratios. Given JWST’s spectral coverage starting around 4~$\mu$m, only select JWST NIRCam data points were used as references for the simulated Twinkle spectra.

To evaluate the retrievability of minor atmospheric species whose spectral signatures are typically obscured by clouds or dominated by major molecules, we conducted a dedicated sensitivity study. Using WASP-107\,b as the test case, we systematically amplified the abundances of minor species (CH$_4$, SO$_2$, NH$_3$, and H$_2$S) to identify the enhancement factor necessary for their spectral features to become clearly observable above simulated observational uncertainties. The baseline scenario employed the previously established retrieval setup for WASP-107\,b, stacking six transits to achieve a median SNR of 10. \rev{For each molecule, the abundance was scaled from its baseline value by amplification factors ranging from $\times$5 to $\times$80 (the specific factors tested for each species are listed in Table~\ref{tab:amplify_lrt}), while all other atmospheric parameters were held fixed at their baseline values. Independent retrieval analyses were then performed at both SNR = 5 and SNR = 10 to determine the minimum amplification factor required for a constrained retrieval of each species.}

\section{Results}
\label{sec:results}

\subsection{Model assumptions: lessons learned from JWST}
With recent advances resulting from JWST observations significantly refining our understanding of exoplanetary atmospheres, previous assumptions regarding atmospheric thickness and composition require reassessment. JWST data have provided critical updates, particularly regarding atmospheric composition and the prevalence of clouds, which directly impact observational strategies and the SNR achievable by Twinkle. 
Table \ref{tab:transit_predictions} summarises our revised estimates of the number of transits required by Twinkle to reach specific SNR targets (3, 5, 7, and 10) at native resolution after incorporating recent JWST findings. We compare these updated estimates with earlier calculations that assumed atmospheric thicknesses of 2, 3, and 5 scale heights. The updated calculations assume ideal observational conditions -- complete transit visibility without interruptions -- thus serving as lower-bound estimates. These preliminary outcomes, integrating JWST observational results, will continue to evolve as additional JWST data become available.

\begin{table*}
    \centering
    \caption{Predicted and realistic number of transits needed for Twinkle to achieve a specific SNR for various planets. Red column indicates the number of transits needed to achieve the updated SNR from JWST, while the Green colour indicates assumptions that is closest to the JWST observation. Real numbers are shown in the brackets and rounded integers are shown next to them. $H$ in the brackets following Predicted Transits is the demonstrate scale height in each planet.}
    \label{tab:transit_predictions}
        \begin{tabular}{cccccc}
        \hline
            \textbf{Planet} & \textbf{Median SNR} & \textbf{Predicted Transits (5$H$)} & \textbf{Predicted Transits (3$H$)} & \textbf{Predicted Transits (2$H$)} & {\color[HTML]{E6000E} \textbf{Realistic Transits}} \\
            \midrule
            \multirow{3}{*}{\textbf{HD\,209458\,b}} 
            & 5  & 1 (0.18) & {\color[HTML]{568F27} 1 (0.51)} & 2 (1.21) & {\color[HTML]{E6000E} 1 (0.85)} \\
            & 7  & 1 (0.35) & {\color[HTML]{568F27} 2 (1.04)} & 3 (2.6)  & {\color[HTML]{E6000E} 2 (1.77)} \\
            & 10 & 1 (0.75) & {\color[HTML]{568F27} 3 (2.32)} & 7 (6.79) & {\color[HTML]{E6000E} 5 (4.24)} \\
            \midrule
            \multirow{3}{*}{\textbf{WASP-39\,b}} 
            & 5  & {\color[HTML]{568F27} 3 (2.26)} & 7 (6.31) & 15 (14.31) & {\color[HTML]{E6000E} 3 (2.11)} \\
            & 7  & {\color[HTML]{568F27} 5 (4.45)} & 13 (12.45) & 29 (28.45) & {\color[HTML]{E6000E} 5 (4.14)} \\
            & 10 & {\color[HTML]{568F27} 10 (9.12)} & 26 (25.74) & 60 (59.84) & {\color[HTML]{E6000E} 9 (8.48)} \\
            \midrule
            \multirow{3}{*}{\textbf{WASP-80\,b}} 
            & 5  & 3 (2.01) & {\color[HTML]{558F28} 6 (5.7)} & 14 (13.41) & {\color[HTML]{E6000E} 8 (7.49)} \\
            & 7  & 4 (3.98) & {\color[HTML]{558F28} 12 (11.56)} & 29 (28.5) & {\color[HTML]{E6000E} 16 (15.35)} \\
            & 10 & 9 (8.33) & {\color[HTML]{558F28} 26 (25.44)} & 71 (70.86) & {\color[HTML]{E6000E} 35 (34.67)} \\
            \midrule
            \multirow{3}{*}{\textbf{WASP-107\,b}} 
            & 5  & 1 (0.16) & 1 (0.45) & {\color[HTML]{568F27} 2 (1.02)} & {\color[HTML]{E6000E} 2 (1.31)} \\
            & 7  & 1 (0.32) & 1 (0.88) & {\color[HTML]{568F27} 2 (2.00)} & {\color[HTML]{E6000E} 3 (2.59)} \\
            & 10 & 1 (0.65) & 2 (1.81) & {\color[HTML]{568F27} 5 (4.15)} & {\color[HTML]{E6000E} 6 (5.4)} \\
            \midrule
            \multirow{3}{*}{\textbf{GJ\,3470\,b}} 
            & 3  & 1 (0.59) & 2 (1.65) & {\color[HTML]{568F27} 4 (3.77)} & {\color[HTML]{E6000E} 4 (3.58)} \\
            & 5  & 2 (1.64) & 5 (4.69) & {\color[HTML]{568F27} 12 (11.09)} & {\color[HTML]{E6000E} 11 (10.49)} \\
            & 7  & 4 (3.27) & 10 (9.55) & {\color[HTML]{568F27} 24 (23.85)} & {\color[HTML]{E6000E} 23 (22.45)} \\
            \bottomrule
        \end{tabular}%
\end{table*}

The realistic observation requirements based on recent JWST constraints are marked in red in Table~\ref{tab:transit_predictions}, while atmospheric thickness predictions closest to the JWST-derived reality are highlighted in green. Initial analyses indicate that Jupiter-sized exoplanets (e.g., HD\,209458\,b, WASP-39\,b, WASP-80\,b) typically have observable atmospheric thicknesses ranging from approximately 2.5 to 5 scale heights. In contrast, Neptune-sized exoplanets (e.g., WASP-107\,b, GJ\,3470\,b) align better with atmospheric thicknesses around two scale heights. These findings broadly confirm previous assumptions but now offer significantly tighter constraints, improving observational efficiency.

Furthermore, JWST data confirm that cloud coverage significantly reduces observable atmospheric thickness. Most H$_2$/He-rich planets studied thus far display observable atmospheric features between 2 and 3 times their scale height. Notably, exceptions such as WASP-39\,b exhibit significantly extended atmospheric signals, a result directly linked to its particularly low planetary density. These refined assumptions provide a robust foundation for the detailed retrieval studies and sensitivity analyses of individual exoplanets presented in subsequent sections.

\subsection{Candidate List for Atmospheric Characterisation Survey}
\label{sec:candidates}

\begin{figure*}
    \centering
    \includegraphics[width=\textwidth]{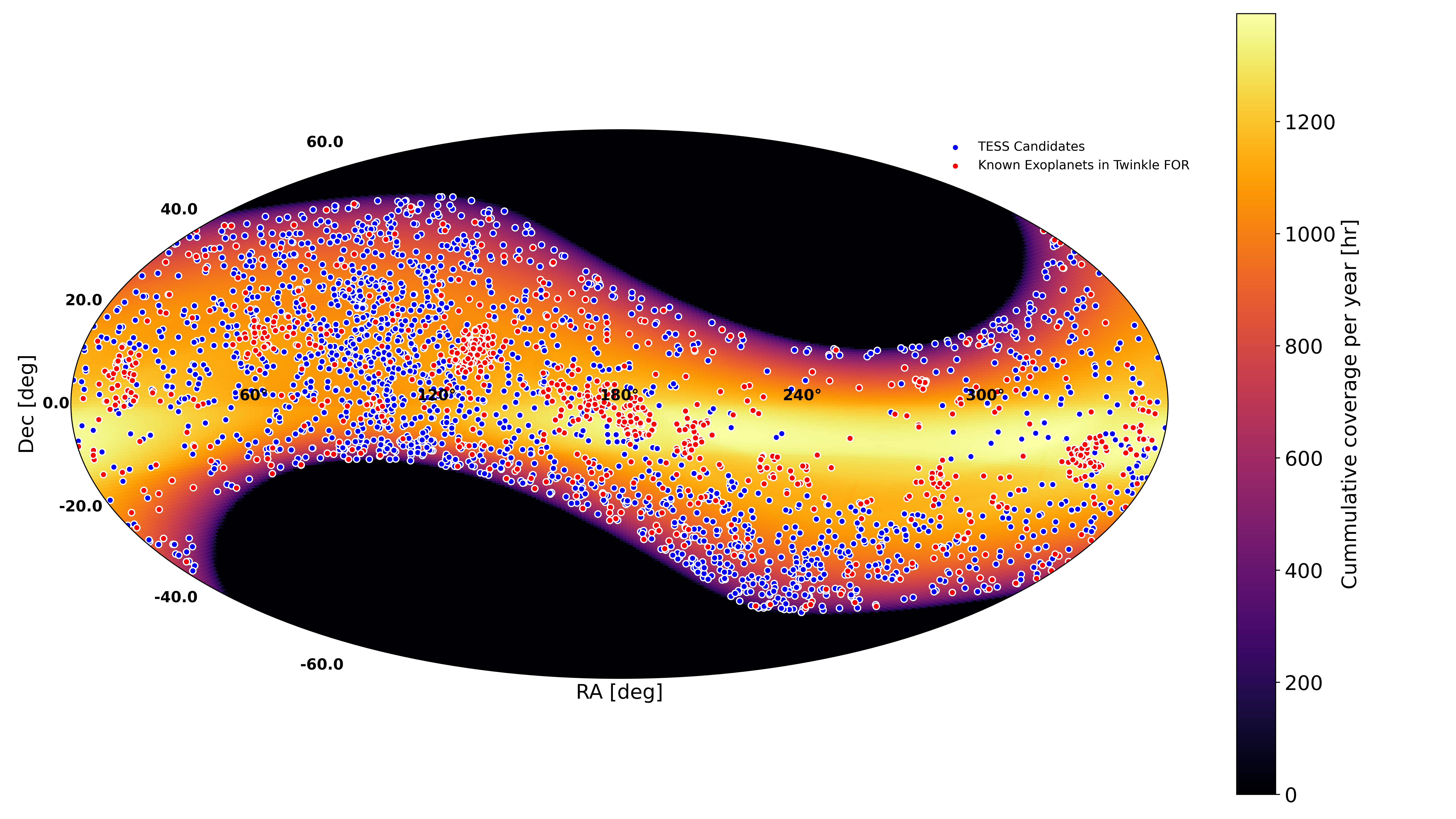} 
    \caption{The sweep of Twinkle's FoR and location of currently known exoplanets (red) and TESS objects of interest (blue).}
    \label{fig:tess+known}
\end{figure*}

Among the confirmed exoplanets, approximately 1,070 transiting exoplanets lie within Twinkle’s FoR. Of the TOIs, 2,009 candidates fall within Twinkle’s FoR. Figure \ref{fig:tess+known} illustrates the right ascension (RA) and declination (Dec) of all confirmed exoplanets and TOIs distribution in the Twinkle’s FoR.

\begin{figure*}
    \centering
    \begin{subfigure}{\textwidth}
        \centering
        \includegraphics[width=\textwidth]{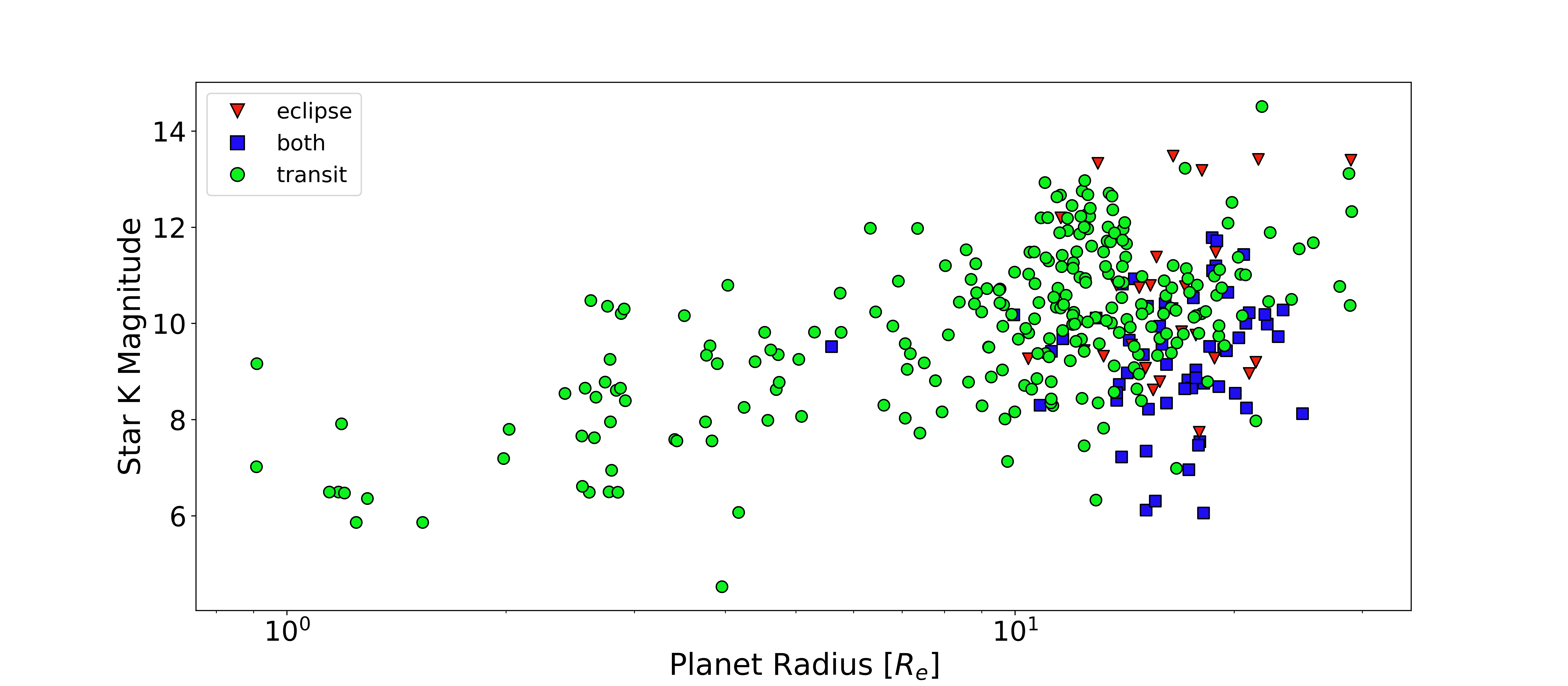}
        \vspace{-4.2em}
    \end{subfigure}
    \begin{subfigure}{\textwidth}
        \centering
        \includegraphics[width=\textwidth]{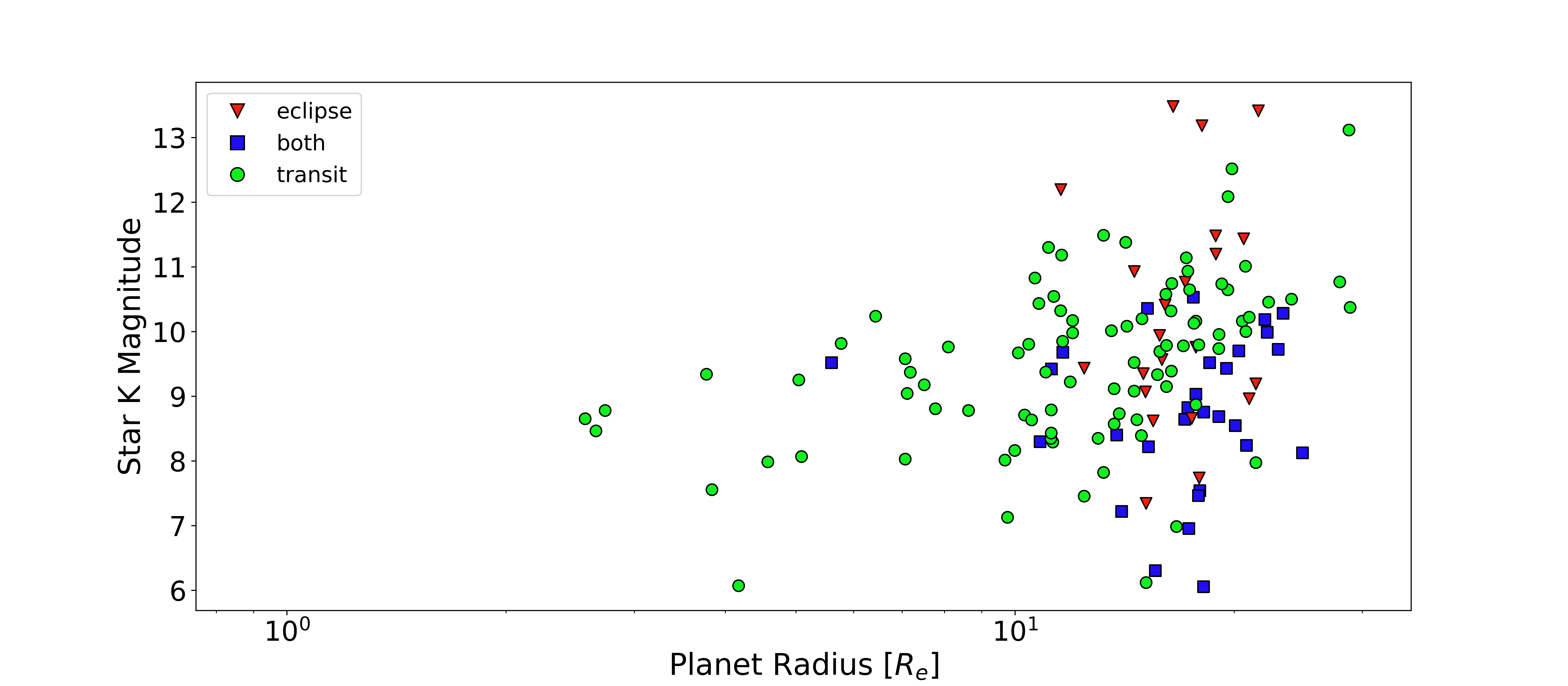}
        \vspace{-4.2em}
    \end{subfigure}
    \begin{subfigure}{\textwidth}
        \centering
        \includegraphics[width=\textwidth]{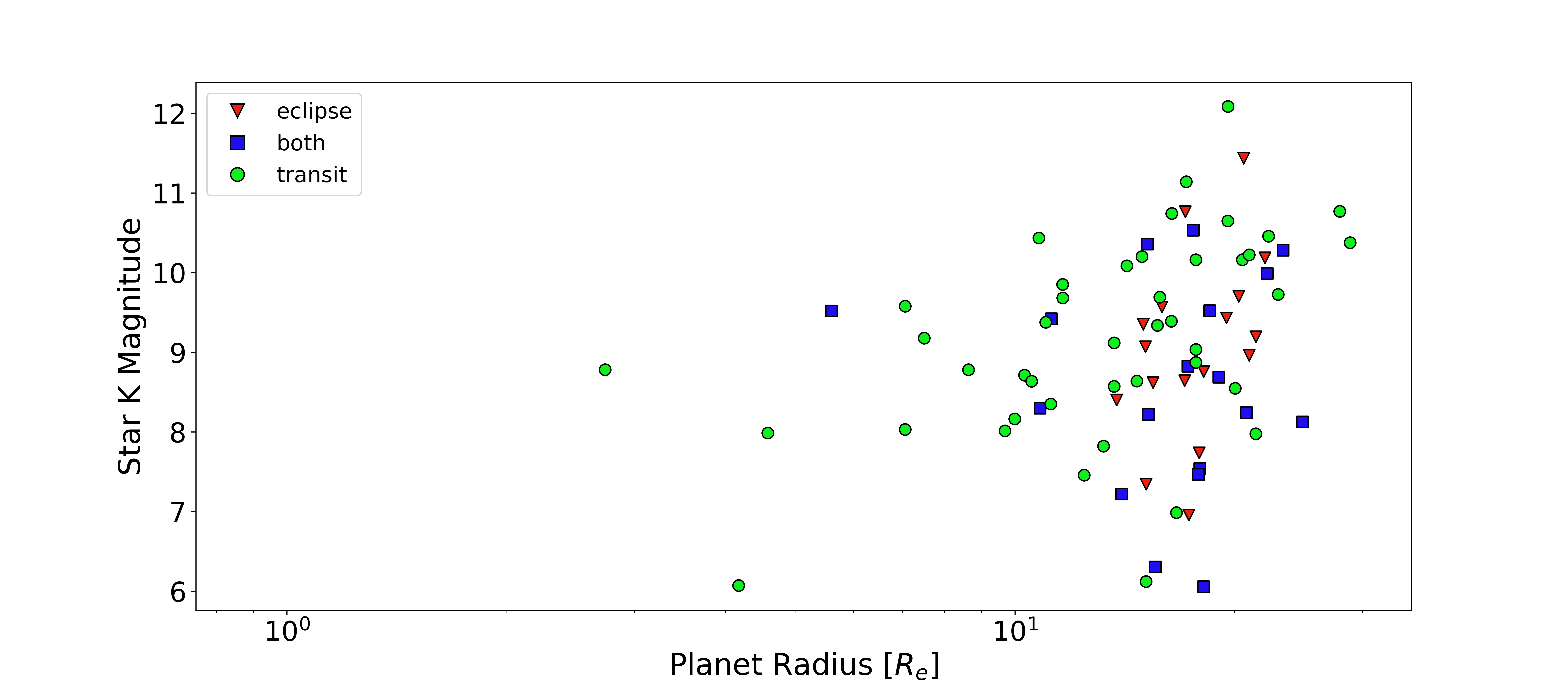}
    \end{subfigure}
    \caption{Known planets and TOIs capable of achieving median SNR thresholds within 30 observations at native resolution. They are distributed based on planet radius versus host star K magnitude. Different shapes demonstrate the preferred type of observation. Top Panel: SNR $\geq$ 3; Middle Panel: SNR $\geq$ 5; Bottom Panel: SNR $\geq$ 7.}
    \label{fig:twinkle_candidates}
\end{figure*}

After processing all targets through the radiometric model, a combined set of known exoplanets and TOIs capable of achieving median SNR thresholds (3, 5, or 7) within 30 observations at native resolution is presented in Figure \ref{fig:twinkle_candidates}.

This initial candidate list used for the observability study for Twinkle’s atmospheric characterisation survey is specifically tailored to H$_2$/He-rich planets, due to current uncertainties surrounding the atmospheric composition of smaller  planets. Recent observations of terrestrial planets have highlighted significant diversity, emphasising the need for further study with dedicated future missions. Consequently, terrestrial planets have not been extensively analysed in this observability study. Further refinement of the candidate list will be informed by complementary observations from JWST, PLATO, CSST, and Earth 2.0, enabling tailored observational strategies aligned with the scientific objectives of the Twinkle Science Team. \rev{The candidate list presented in this work will be made publicly available to support the broader exoplanet community in planning complementary observations.}

\subsection{Spectral Retrievals: Gas Giants}
\subsubsection{HD\,209458\,b}
\label{sec:hd209}

\begin{figure*}
    \centering
    \includegraphics[width=\textwidth]{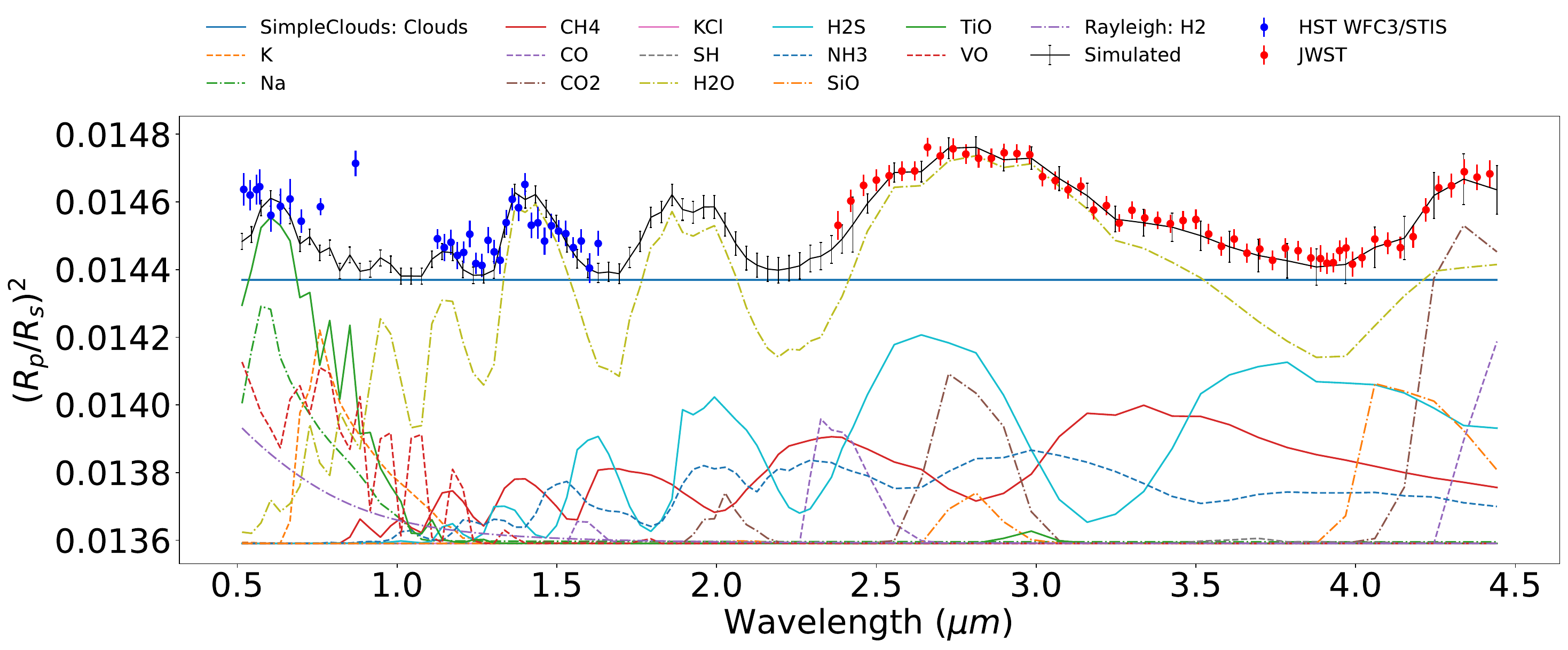} 
    \caption{Comparison of the JWST and Twinkle-simulated spectrum for HD\,209458\,b, highlighting the spectral contributions of various molecular absorbers and scattering effects. The error bars on the spectrum are based on simulations of 10 transits, and the spectra are fitted under the assumption of chemical equilibrium. For clarity, spectral contributions from effects that are insignificant have been omitted. The HST data points have been shifted downward by 100 ppm.}
    \label{fig:hd209_RCE_sim}
\end{figure*}

\begin{figure}
    \centering
    \includegraphics[width=\columnwidth]{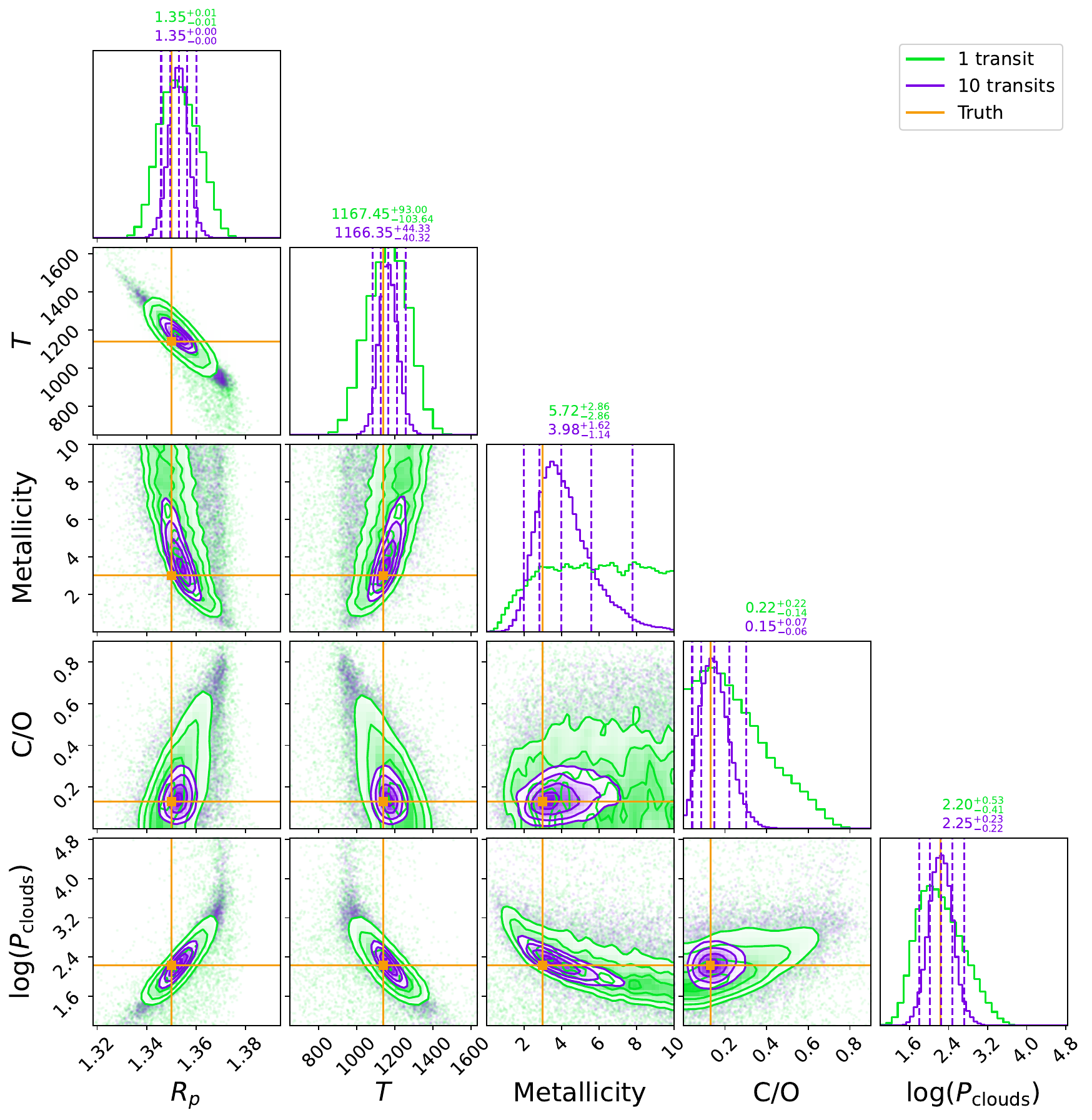}
    \caption{Posterior distributions of retrieved parameters from retrieval analysis of HD\,209458\,b assuming chemical equilibrium. Green: 1 transit; Purple: 10 transits.}
    \label{fig:RCE_post}
\end{figure}

Simulated spectra for HD\,209458\,b with Twinkle's predicted uncertainties after fitting with combined set of HST WFC3 and JWST are shown in Figure~\ref{fig:hd209_RCE_sim}. Retrievals conducted under chemical equilibrium assumptions indicate that, even with a single transit, abundances of major molecules can be accurately recovered. However, minor molecular species such as CH\textsubscript{4}, SO\textsubscript{2}, and HCN exhibit weaker constraints, introducing uncertainties in metallicity and C/O ratio estimates (Figure \ref{fig:RCE_post}). Increasing observations to 10 transits significantly enhances the accuracy of CH\textsubscript{4} and HCN abundance constraints, leading to more robust metallicity and C/O ratio estimates (Figure \ref{fig:RCE_post}). This improvement mainly results from the distinct and abundant H\textsubscript{2}O absorption due to the assumed low C/O ratio, prominent TiO features at optical wavelengths, and clearly defined CO\textsubscript{2} features at the  limit of Twinkle’s coverage. Additional planetary parameters, including radius, atmospheric temperature, and cloud-top pressure, were also accurately recovered within $1\sigma$.

To provide a more comprehensive assessment, we performed a free retrieval analysis without assuming chemical equilibrium by selecting the eleven most critical molecules identified in previous simulations, along with species suggested by \citet{2024ApJ...963L...5X}. A constant vertical mixing ratio (VMR) profile based on equilibrium abundances and suggestions from \citet{2024ApJ...963L...5X} was used for these molecules in a free retrieval analysis. Results in Figure~\ref{fig:hd209_free_post} confirm robust retrieval of H\textsubscript{2}O \rev{and} TiO, whereas other molecular signatures remain masked by the dominant H\textsubscript{2}O absorption and cloud opacity. Twinkle's extensive spectral coverage from optical to infrared allows for clear identification of TiO \rev{absorption features, while VO remains below the detection threshold (LRT $\sim 2\sigma$; Table~\ref{tab:hd209_retrieved_para}) despite showing a posterior shift from the prior}. However, due to the relatively low assumed abundance of CO\textsubscript{2} ($10^{-6}$) and increased noise at longer wavelengths, the CO\textsubscript{2} feature was only partially detected and thus less reliably constrained.

Retrieval comparisons at different SNR values (5, 7, and 10) for HD\,209458\,b, as illustrated in Figure \ref{fig:hd209_free_post}, indicate improved parameter constraints for strongly featured molecules such as H\textsubscript{2}O and TiO at higher SNR. Conversely, parameters not adequately retrieved at an SNR of 5 exhibit limited improvement at higher SNR levels. This limitation of improving SNR and difficulty of retrieving minor species in HD\,209458\,b likely arises from spectral masking by prominent water absorption and simplified grey cloud assumptions, rather than reflecting Twinkle's intrinsic capabilities. 

\begin{table}
\renewcommand{\arraystretch}{1.3}
\centering
\caption{Comparison of truth values and retrieved parameters for HD\,209458\,b with 1 transit (SNR = 5).}
\label{tab:hd209_retrieved_para}
\begin{tabular}{lccc}
\specialrule{1pt}{1pt}{1pt}
\textbf{Parameter} & \textbf{Truth} & \textbf{Retrieved (SNR = 5)} & \rev{\textbf{LRT ($\sigma$)}} \\
\midrule
Planet Radius [$R_J$] & 1.35 & $1.34^{+0.01}_{-0.01}$ \\
Temperature [K] & 1140 & $1205^{+118}_{-122}$ \\
\midrule
\multicolumn{3}{l}{\textbf{Detection}} \\
$\log(X_{\mathrm{H_2O}})$ & $-2.7$ & $-2.25^{+0.57}_{-0.62}$ & \rev{15.18} \\
$\log(X_{\mathrm{TiO}})$ & $-8$ & $-7.51^{+0.61}_{-0.63}$ & \rev{8.99}\\
\midrule
\multicolumn{3}{l}{\textbf{Upper Limit}} \\
$\log(X_{\mathrm{VO}})$ & $-9$ & $-9.31^{+0.93}_{-1.42}$ & \rev{2.05}\\
$\log(X_{\mathrm{CO_2}})$ & $-6$ & $-7.63^{+2.20}_{-2.80}$ & \rev{1.20}\\
$\log(X_{\mathrm{CO}})$ & $-3.4$ & $-7.39^{+3.33}_{-2.96}$ & \rev{$<0.01$}\\
$\log(X_{\mathrm{H_2S}})$ & $-5$ & $-7.57^{+2.78}_{-2.82}$ & \rev{$<0.01$}\\
$\log(X_{\mathrm{NH_3}})$ & $-5.3$ & $-7.99^{+2.47}_{-2.60}$ & \rev{$<0.01$}\\
$\log(X_{\mathrm{HCN}})$ & $-6$ & $-8.28^{+2.52}_{-2.44}$ & \rev{0.10}\\
$\log(X_{\mathrm{CH_4}})$ & $-8$ & $-8.57^{+2.25}_{-2.20}$ & \rev{0.05}\\
$\log(X_{\mathrm{C_2H_2}})$ & $-8$ & $-8.30^{+2.25}_{-2.34}$ & \rev{0.12}\\
\midrule
$\log(P_{\text{clouds}} [Pa])$ & 2.11 & $1.63^{+0.59}_{-0.53}$ \\
\bottomrule
\end{tabular}
\end{table}

\subsubsection{WASP-107\,b}
\label{sec:wasp107b}

\begin{figure*}
    \centering
    \includegraphics[width=\textwidth]{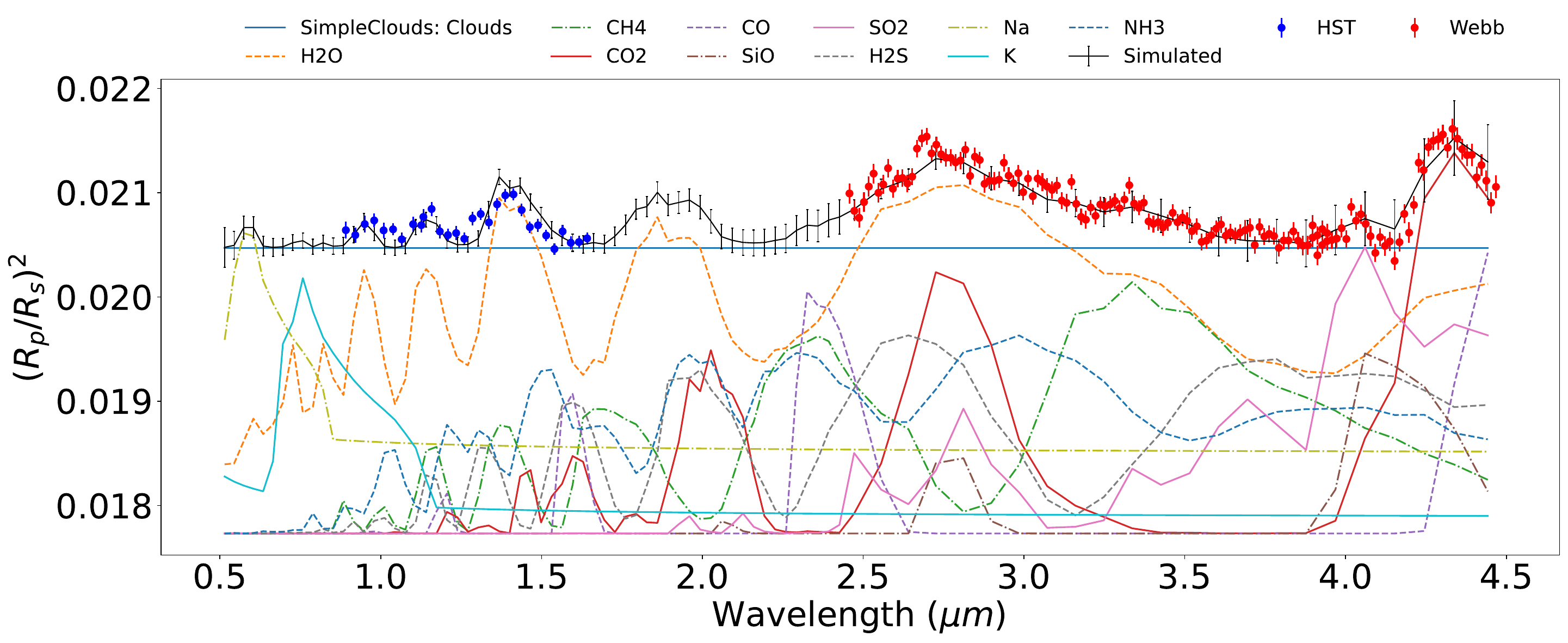}
    \caption{Comparison of JWST transmission spectrum and Twinkle-simulated spectrum, with spectral contributions of various molecule absorbers and scattering effects for WASP-107\,b. Error bar on spectrum is simulated with observation of 6 transits (SNR = 10). The spectra is fitted assuming free chemistry. The JWST data points were offset upward by 300 ppm.}
    \label{fig:wasp107b_jwst_vs_simulated}
\end{figure*}

Simulated spectrum for WASP-107\,b with Twinkle's predicted error after fitting with available HST WFC3 and JWST NIRCam from \citet{2024Natur.630..836W} are shown in Figure~\ref{fig:wasp107b_jwst_vs_simulated}. Calculations shown in Table \ref{tab:transit_predictions} suggest the atmospheric thickness of WASP-107\,b is slightly less than two scale heights, likely due to high-altitude cloud cover. To reflect this observational constraint, our simulation adopted a high cloud altitude at 70 Pa.

Posterior distributions for WASP-107\,b in Figure \ref{fig:wasp107b_free_retrieval_results} revealed significant degeneracies between the planetary radius and cloud-top height parameters, with the radius typically underestimated and cloud height overestimated in retrievals. Nevertheless, robust constraints were obtained for the equilibrium temperature and the abundances of H\textsubscript{2}O, CO\textsubscript{2}, and Na. The abundances of NH\textsubscript{3} and CH\textsubscript{4} remainundetectable due to their lower contribution compared with cloud opacity within Twinkle’s spectral coverage as in Figure \ref{fig:wasp107b_jwst_vs_simulated}. The primary absorption band for CO (4.5–5.0~$\mu$m) lies outside Twinkle's spectral range, making CO undetectable. The spectral signature of SO\textsubscript{2} at 4.0~$\mu$m is too weak to be reliably retrieved, primarily due to interference from cloud opacity and observational noise.

Comparative retrieval results at different SNR values showed that parameters clearly observable in the spectra, such as H\textsubscript{2}O and CO\textsubscript{2}, exhibited tighter constraints with increasing SNR. Parameters initially unconstrained at an SNR of 5 but exhibiting spectral contributions stronger than cloud opacity—such as Na—became retrievable at higher SNR values (SNR = 10). Furthermore, while CO\textsubscript{2} could be constrained at all simulated SNR levels, increasing the SNR to 7 or 10 significantly improved the accuracy and reliability of its retrieval.

\begin{table}
\renewcommand{\arraystretch}{1.3}
\centering
\normalsize
\caption{Comparison of truth values and retrieved parameters (RP) for WASP-107\,b stacking 2 transits (SNR = 5).}
\label{tab:wasp107b_retrieved_para}
\begin{tabular}{lccc}
\specialrule{1pt}{1pt}{1pt}
\textbf{Parameter} & \textbf{Truth} & \textbf{RP (SNR = 5)} & \rev{\textbf{LRT ($\sigma$)}} \\
\midrule
Planet Radius [$R_J$] & 0.933 & $0.90 \pm 0.02$ \\
Temperature [K] & 700 & $689^{+91}_{-88}$ \\
\midrule
\multicolumn{3}{l}{\textbf{Detection}} \\
$\log(X_{\mathrm{H_2O}})$ & $-2$ & $-2.07^{+0.51}_{-0.64}$ & \rev{13.39}\\
\midrule
\multicolumn{3}{l}{\textbf{Upper Limit}} \\
$\log(X_{\mathrm{CO}})$ & $-1.4$ & $-7.35^{+3.08}_{-2.81}$ & \rev{$<0.01$}\\
$\log(X_{\mathrm{CO_2}})$ & $-4.52$ & $-5.30^{+1.17}_{-2.52}$ & \rev{1.35}\\
$\log(X_{\mathrm{Na}})$ & $-4$ & $-7.03^{+2.65}_{-2.92}$ & \rev{0.07} \\
$\log(X_{\mathrm{CH_4}})$ & $-5.7$ & $-8.35^{+2.12}_{-2.24}$ & \rev{$<0.01$}\\
$\log(X_{\mathrm{H_2S}})$ & $-4.3$ & $-7.62^{+2.70}_{-2.69}$ & \rev{$<0.01$}\\
$\log(X_{\mathrm{K}})$ & $-6$ & $-8.46^{+2.31}_{-2.19}$ & \rev{0.08}\\
$\log(X_{\mathrm{NH_3}})$ & $-5.7$ & $-8.09^{+2.25}_{-2.41}$ & \rev{$<0.01$}\\
$\log(X_{\mathrm{SO_2}})$ & $-5$ & $-8.13^{+2.47}_{-2.32}$ & \rev{$<0.01$}\\
$\log(X_{\mathrm{SiO}})$ & $-4$ & $-7.18^{+3.21}_{-3.05}$ & \rev{0.01}\\
\midrule
$\log(P_{\text{clouds}} [Pa])$ & $1.85$ & $1.49^{+0.62}_{-0.47}$ \\
\bottomrule
\end{tabular}
\end{table}

\subsubsection{GJ\,3470\,b}
\label{sec:gj3470b}

\begin{figure*}
    \centering
    \includegraphics[width=\textwidth]{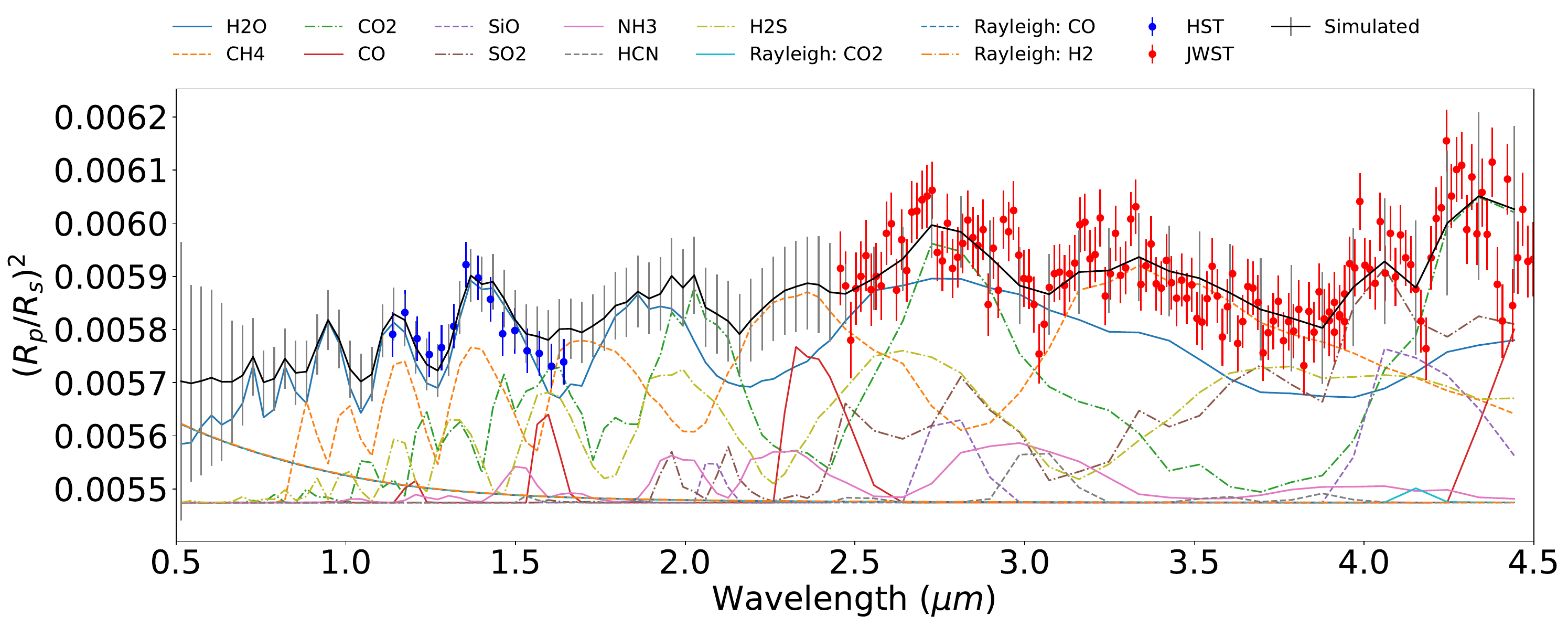}
    \caption{Comparison of JWST and Twinkle-simulated spectrum and spectral contributions of various molecule absorbers and scattering effects for GJ\,3470\,b. Error bar on spectrum is simulated with observation of 11 transits (SNR = 5). The spectra is fitted assuming free chemistry. The HST data points were offset downward by 200 ppm. }
    \label{fig:gj3470b_jwst_vs_simulated}
\end{figure*}

Simulated spectra of GJ\,3470\,b with Twinkle's predicted error after fitting with available HST WFC3 and JWST NIRCam from \citet{2025ApJS..276...70S} and \citet{2024ApJ...970L..10B} are shown in Figure~\ref{fig:gj3470b_jwst_vs_simulated}. The contribution plot illustrates distinct spectral features from key molecules. In addition to strong absorption by H\textsubscript{2}O, significant CO\textsubscript{2} features at 2.0, 3.8, and 4.4~$\mu$m allow robust constraints on its abundance. Prominent CH\textsubscript{4} absorption features at 1.6, 2.4, and 3.4~$\mu$m are also clearly discernible. A detectable but weaker absorption peak of SO\textsubscript{2} at 4.0~$\mu$m provides crucial insight into atmospheric disequilibrium chemistry.

Considering the observational challenges posed by GJ\,3470\,b's low transit depth and relatively flat spectrum, we evaluated retrieval outcomes at SNR levels of 3, 5, and 7, approximately 23 transits are required to reach an SNR of 7. The retrieval results, depicted in Figure \ref{fig:gj3470b_free_retrieval_results}, demonstrate accurate constraints for parameters such as planetary radius and equilibrium temperature. Conversely, the cloud deck height remains unconstrained due to its depth, rendering it indistinguishable from other atmospheric components. The abundances of H\textsubscript{2}O, CH\textsubscript{4}, and CO\textsubscript{2} were successfully retrieved at all SNRs considered. However, SO\textsubscript{2} could only be effectively constrained at the highest SNR (SNR = 7). Minor molecular species such as CO, SiO, NH\textsubscript{3}, HCN, and H\textsubscript{2}S were undetectable even at high SNR due to their weak absorption features within Twinkle's spectral range.

Comparative retrieval results across different SNR scenarios yielded consistent conclusions with previous exoplanet cases. Parameters with strong spectral signatures above cloud opacity, such as H\textsubscript{2}O, CO\textsubscript{2}, and CH\textsubscript{4}, benefit significantly from higher SNR, resulting in tighter constraints. Conversely, molecules like SO\textsubscript{2}, with a single prominent absorption feature near the spectral range boundary, require high SNR levels for accurate abundance retrieval.

\begin{table}
\renewcommand{\arraystretch}{1.3}
\centering
\normalsize
\caption{Comparison of truth values and retrieved parameters (RP) for GJ\,3470\,b stacking 4 transits (SNR = 3).}
\label{tab:gj3470b_retrieved_para}
\begin{tabular}{lccc}
\specialrule{1pt}{1pt}{1pt}
\textbf{Parameter} & \textbf{Truth} & \textbf{RP (SNR = 3)} & \rev{\textbf{LRT ($\sigma$)}}\\
\midrule
Planet Radius [$R_J$] & 0.36 & $0.36 \pm 0.002$ \\
Temperature [K] & 600 & $650^{+235}_{-192}$ \\
\midrule
\multicolumn{3}{l}{\textbf{Detection}} \\
$\log(X_{\mathrm{H_2O}})$ & $-1$ & $-0.92^{+0.51}_{-0.89}$ & \rev{5.03}\\
$\log(X_{\mathrm{CH_4}})$ & $-3.3$ & $-3.48^{+0.80}_{-1.75}$ & \rev{3.37}\\
$\log(X_{\mathrm{CO_2}})$ & $-1.3$ & $-1.48^{+0.66}_{-2.98}$ & \rev{9.52}\\
\midrule
\multicolumn{3}{l}{\textbf{Upper Limit}} \\
$\log(X_{\mathrm{SiO}})$ & $-2$ & $-4.62^{+3.29}_{-4.25}$ & \rev{0.02}\\
$\log(X_{\mathrm{SO_2}})$ & $-3.3$ & $-3.84^{+2.44}_{-4.35}$ & \rev{0.39}\\
$\log(X_{\mathrm{CO}})$ & $-1$ & $-5.03^{+3.48}_{-3.96}$ & \rev{$<0.01$}\\
$\log(X_{\mathrm{NH_3}})$ & $-7$ & $-7.46^{+2.61}_{-2.59}$ & \rev{$<0.01$}\\
$\log(X_{\mathrm{HCN}})$ & $-8$ & $-5.91^{+2.85}_{-3.50}$ & \rev{$<0.01$}\\
$\log(X_{\mathrm{H_2S}})$ & $-3$ & $-5.71^{+3.20}_{-3.43}$ & \rev{$<0.01$}\\
\midrule
$\log(P_{\text{clouds}} [Pa])$ & $3.48$ & $4.27^{+0.95}_{-0.99}$ \\
\bottomrule
\end{tabular}
\end{table}

\subsection{Spectral Retrievals: super-Earths}
\subsubsection{55\,Cancri\,e}
\label{sec:55cnce}

\begin{figure*}
    \centering
    \includegraphics[width=\textwidth]{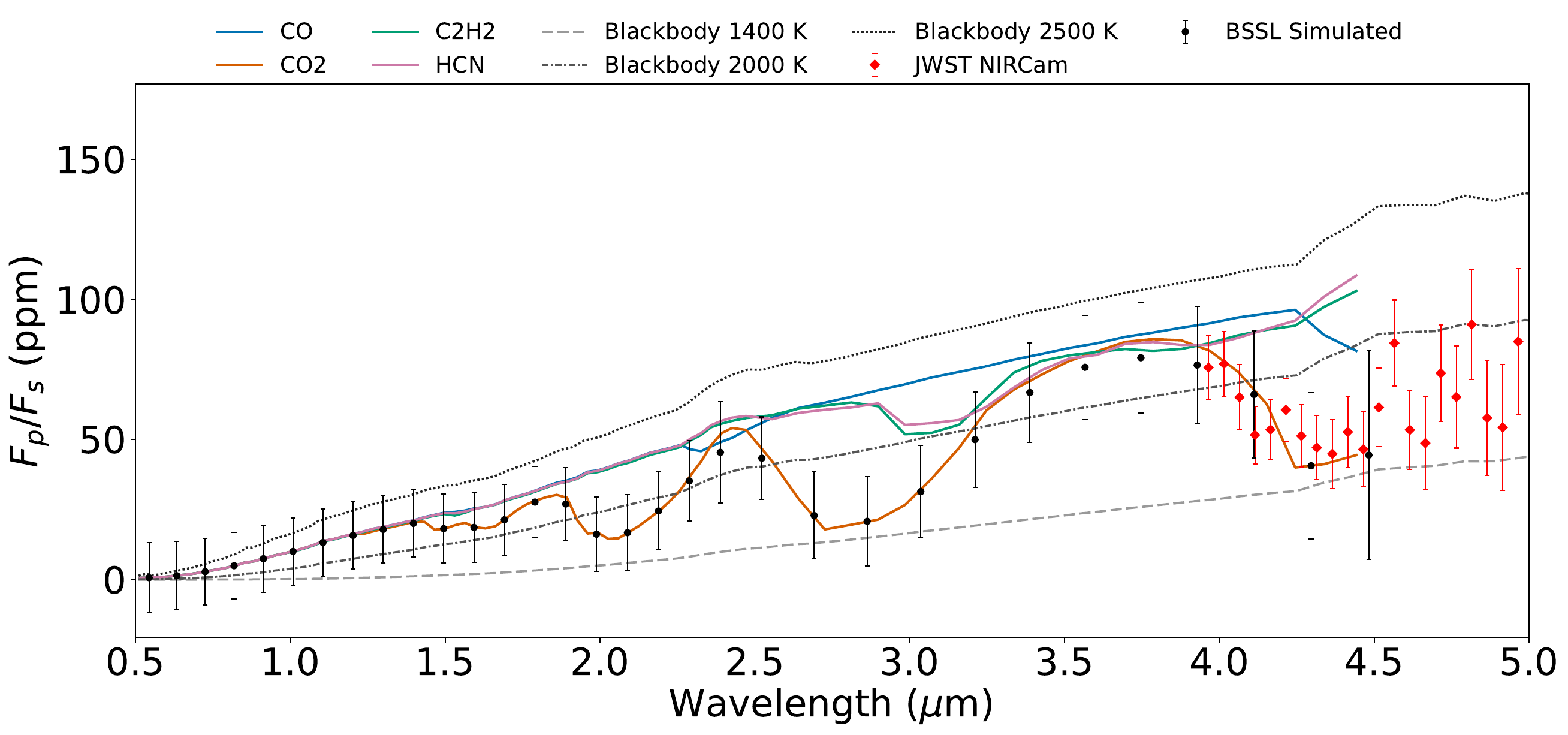}
    \caption{Simulated emission spectrum of 55\,Cnc\,e (black circles) with spectral contributions of individual molecular absorbers (coloured lines), three blackbody reference curves at 1400, 2000, and 2500\,K (grey lines), and JWST NIRCam observations (red diamonds; \citealt{2024Natur.630..609H}). Error bars on the simulated data are for 10 eclipses with ch0 binned to one-third and ch1 binned to half of the native resolution; native resolution with 10 eclipses was used in retrieval.}
    \label{fig:55cnce_BB}
\end{figure*}

Figure~\ref{fig:55cnce_BB} shows a simulated spectrum for 55\,Cnc\,e, along with Twinkle’s predicted error after fitting with available JWST NIRCam data from \citet{2024Natur.630..609H}. This baseline scenario uses three blackbody curves representing temperatures of 1400, 2000, and 2500 K. Comparing these baselines with the simulated and observed data points reveals distinctive absorption features present in the emission spectrum. Twinkle’s expected performance, calculated by reducing the resolution of channel 0 to one-third and channel 1 to half their native resolutions and stacking ten eclipse observations, demonstrates comparable sensitivity to JWST within its spectral range. Figure~\ref{fig:55cnce_BB} reveals significant spectral divergence from blackbody emission beginning around 2.0~$\mu$m. Although the higher resolution of channel 0 below this wavelength is underutilised, the spectral range above 2.0~$\mu$m effectively captures prominent CO\textsubscript{2} absorption features. Additional CO\textsubscript{2} features are clearly identifiable within Twinkle’s channel 1 coverage. The contribution curves further highlight that CO\textsubscript{2} dominates the spectral signature throughout Twinkle's wavelength range, overshadowing contributions from minor species such as C\textsubscript{2}H\textsubscript{2}, CO, and HCN. Notably, an increase in CO contribution is evident beyond 4.2~$\mu$m, suggesting enhanced detectability if Twinkle’s coverage could extend slightly further. 

Using this simulated spectrum as input, we performed emission retrievals to evaluate Twinkle’s capability in constraining atmospheric properties. To \rev{avoid the loss of spectral information introduced by post-processing wavelength binning}, retrievals were conducted at Twinkle's native resolution, despite larger resultant uncertainties. Retrieval outcomes for scenarios involving 10, 20, and 30 stacked eclipse observations are presented in Figure \ref{fig:55cnce_retri}. The results illustrate that, with just 10 eclipses, Twinkle effectively constrains the temperature structure and reliably retrieves the abundance of the dominant atmospheric species, CO\textsubscript{2}. Additional observations enhance parameter constraints incrementally but do not significantly improve the detection of dominant atmospheric constituents. This demonstrates Twinkle's robust performance, particularly for bright \rev{super-Earth targets such as} 55\,Cnc\,e.

\begin{table}
\renewcommand{\arraystretch}{1.3}
\centering
\caption{Comparison of truth values and retrieved parameters (RP) for 55\,Cnc\,e stacking 10 eclipses.}
\label{tab:55cnce_retrieved_para}
\begin{tabular}{lccc}
\specialrule{1pt}{1pt}{1pt}
\textbf{Parameter} & \textbf{Truth} & \textbf{RP (10 eclipses)} & \rev{\textbf{LRT ($\sigma$)}}\\
\midrule
Planet Radius [$R_\mathrm{J}$] & 0.167 & $0.150^{+0.01}_{-0.02}$ \\
$T_{\text{surface}}$ [K] & 2300 & $2487^{+204}_{-169}$ \\
$T_{\text{top}}$ [K] & 1500 & $1585^{+232}_{-273}$ \\
$\log(P_{\text{surface}}$ [Pa]) & 5 & $5.57^{+1.50}_{-1.01}$ \\
$\log(P_{\text{top}}$ [Pa]) & 2 & $1.56^{+1.29}_{-0.98}$ \\
\midrule
\multicolumn{3}{l}{\textbf{Detection}} \\
$\log(X_{\mathrm{CO_2}})$ & $-2$ & $-2.38^{+1.01}_{-1.19}$ & \rev{5.38}\\
\midrule
\multicolumn{3}{l}{\textbf{Upper Limit}} \\
$\log(X_{\mathrm{CO}})$ & $-3$ & $-6.32^{+3.34}_{-3.43}$ & \rev{0.01}\\
$\log(X_{\mathrm{HCN}})$ & $-5$ & $-7.92^{+2.49}_{-2.53}$ & \rev{0.01}\\
$\log(X_{\mathrm{C_2H_2}})$ & $-5$ & $-8.17^{+2.53}_{-2.47}$ & \rev{$<0.01$}\\
\bottomrule
\end{tabular}
\end{table}

\subsection{Minor Species Sensitivity Study on WASP-107 b}
\label{sec:minor_species}
Contribution plots from simulations (Figures \ref{fig:wasp107b_jwst_vs_simulated}, \ref{fig:gj3470b_jwst_vs_simulated}, and \ref{fig:55cnce_BB}) revealed that several minor species, such as CH$_4$, SO$_2$, NH$_3$, and H$_2$S, typically possess abundances too low to yield distinguishable spectral features, especially in atmospheres dominated by water vapour absorption or obscured by clouds. Direct retrieval of these species from featureless continuum regions was thus found unreliable, as illustrated in posterior plots provided in the Appendix. Such limitations reflect realistic observational challenges faced when attempting to detect minor constituents even at higher SNR levels.

\begin{figure*}
    \centering
    \includegraphics[width=\textwidth]{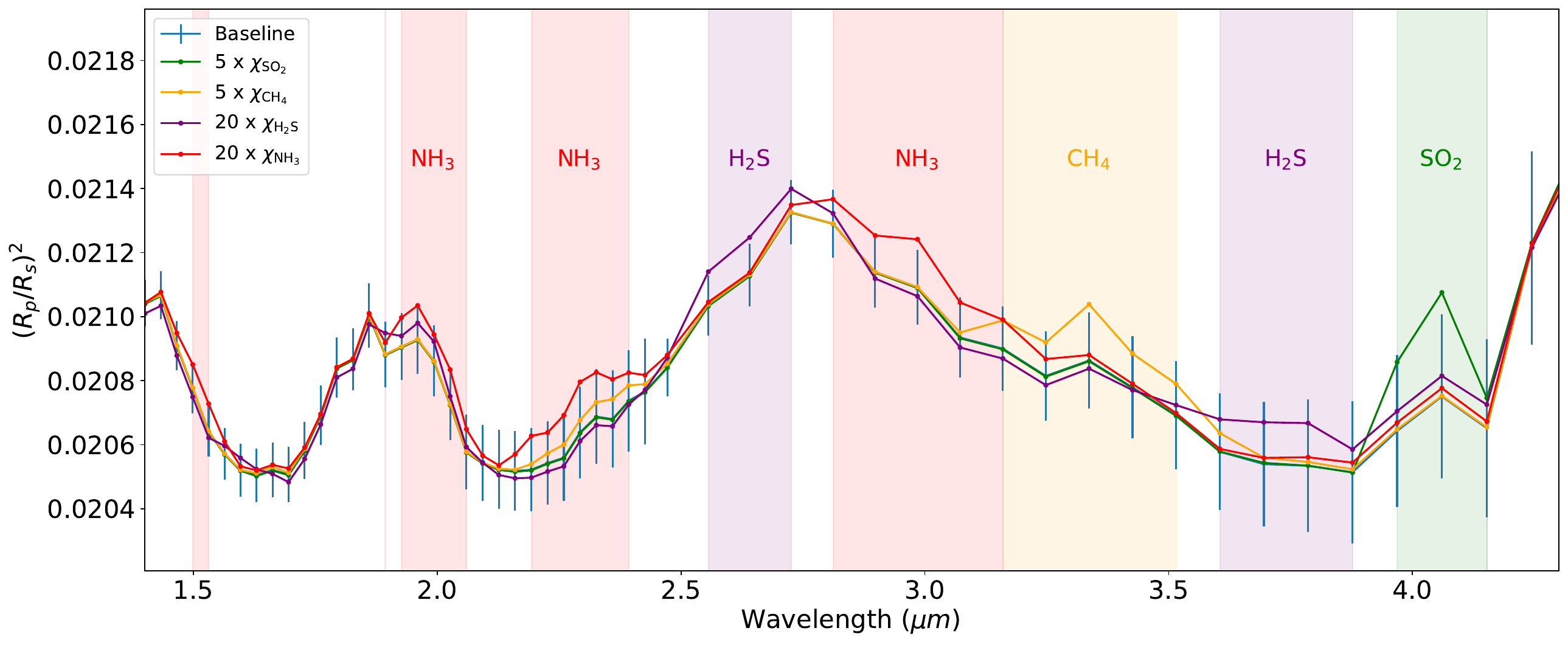}
    \caption{The resultant spectrum with amplifying abundance for CH$_4$, SO$_2$, NH$_3$ and H$_2$S at SNR = 10. Magnifying factor are shown in the legend, assuring visibility in the spectrum. Errorbar on baseline spectrum is estimated with observation of 6 transits. The wavelength ranges where the molecular spectra contribute significantly after abundance amplification have been shaded for visual ease.}
    \label{fig:amplify_spec}
\end{figure*}

To better quantify these observational constraints, we assessed the amplification factors necessary for each minor species’ spectral features to surpass the noise threshold of Twinkle observations at SNR = 10 (Comparison of the original and amplified spectra in Figure~\ref{fig:amplify_spec}). Our results show that CH$_4$ and SO$_2$ features become clearly observable when their abundances increase by approximately fivefold, corresponding to abundances of $1\times10^{-5}$ and $5\times10^{-5}$, respectively. For NH$_3$ and H$_2$S, which have broader and weaker absorption bands overlapping significantly with water vapour, spectral signatures required approximately a twentyfold abundance increase (to $4\times10^{-5}$ and $1\times10^{-3}$) before becoming discernible.

\begin{figure*}
    \centering
    \includegraphics[width=0.8\textwidth]{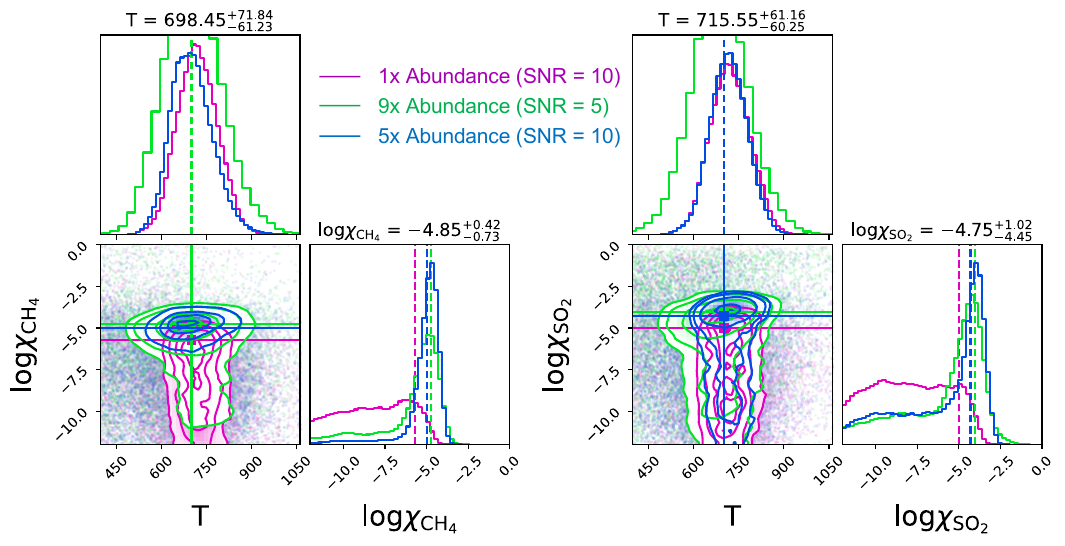}
    \vspace{1cm}
    \includegraphics[width=0.8\textwidth]{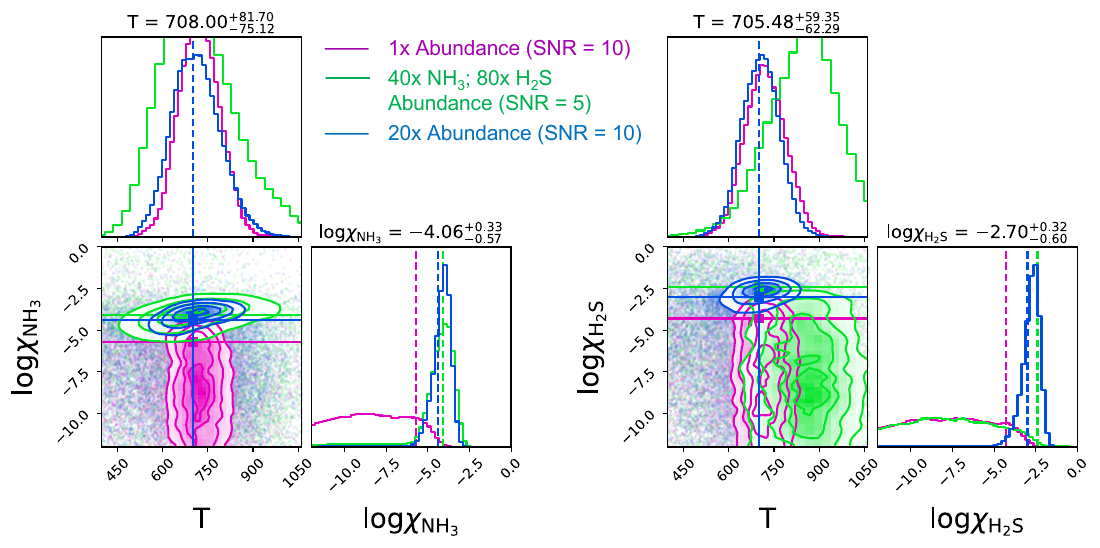}
    \caption{Posteriors for retrievals with amplified abundance for CH$_4$, SO$_2$, NH$_3$ and H$_2$S at different SNR levels for WASP-107\,b. Baseline (Pink) is same with SNR = 10 case in Figure \ref{fig:wasp107b_free_retrieval_results}. Magnifying factor are shown in the legend, assuring visibility in the spectrum.}
    \label{fig:amplify_post}
\end{figure*}

Individual retrieval analyses for each amplified abundance scenario confirmed that minor species, initially undetectable at baseline abundances, can indeed be reliably retrieved when their spectral signatures exceed noise thresholds (posterior distributions in Figure~\ref{fig:amplify_post}). At Twinkle’s native resolution with SNR = 10, CH$_4$ and SO$_2$ could be constrained down to abundances of $1\times10^{-5}$ and $5\times10^{-5}$, respectively. Lowering the SNR to 5 increased these retrieval thresholds to approximately $2\times10^{-5}$ for CH$_4$ and $9\times10^{-5}$ for SO$_2$. NH$_3$ could be reliably constrained at an abundance of $4\times10^{-5}$ at SNR = 10, requiring further amplification to $8\times10^{-5}$ (fortyfold increase from baseline) at SNR = 5. H$_2$S retrieval was challenging at SNR = 5 due to its weaker, indistinct absorption features compared to water.

\rev{While the amplified spectra may appear visually consistent with the baseline within individual spectral bin error bars (Figure~\ref{fig:amplify_spec}), the LRT analysis captures the cumulative effect of systematic spectral differences across multiple bins simultaneously. The retrieval posteriors in Figure~\ref{fig:amplify_post} confirm that the amplified species are indeed constrained, validating the sensitivity thresholds identified by our analysis.}

\begin{table}
\caption{\rev{LRT detection significance ($\sigma$) for minor species with amplified abundances in WASP-107\,b. Values exceeding the $3\sigma$ detection threshold are shown in bold. ``--'' indicates no retrieval was performed for that combination.}}
\label{tab:amplify_lrt}
\centering
\begin{tabular}{llcc}
\toprule
Molecule & Amplification & SNR\,10 & SNR\,5 \\
\midrule
SO$_2$  & baseline       & $<0.01$ & $<0.01$ \\
        & $\times$5      & \textbf{4.69} & 0.12 \\
        & $\times$9      & --    & \textbf{6.20} \\
\midrule
NH$_3$  & baseline       & 0.01 & --    \\
        & $\times$5      & 0.03 & --    \\
        & $\times$10     & \textbf{8.23} & --    \\
        & $\times$20     & \textbf{13.58} & --    \\
        & $\times$40     & --    & \textbf{11.99} \\
        & $\times$50     & \textbf{24.10} & --    \\
\midrule
H$_2$S  & baseline       & $<0.01$ & --    \\
        & $\times$5      & 0.17 & --    \\
        & $\times$10     & 2.42 & --    \\
        & $\times$20     & \textbf{3.87} & --    \\
        & $\times$50     & \textbf{24.44} & --    \\
        & $\times$80     & --    & $<0.01$ \\
\bottomrule
\end{tabular}
\end{table}

\rev{Table~\ref{tab:amplify_lrt} reports the LRT detection significance for each amplification scenario. SO$_2$ achieves robust detection ($>3\sigma$) at $\times$5 amplification with SNR\,10 and at $\times$9 amplification ($6.20\sigma$) with SNR\,5. NH$_3$ requires $\geq\times$10 amplification at SNR\,10, while H$_2$S requires $\geq\times$20. CH$_4$ is not included in the table because it remains undetected by the relative LRT (full model versus model without CH$_4$) at all tested amplification levels. This non-detection arises from strong spectral degeneracy with H$_2$O, whose retrieved abundance (log\,VMR\,$\approx -1.8$) produces broad near-infrared absorption that overlaps and masks the CH$_4$ features; the retriever can compensate for the removal of CH$_4$ by adjusting H$_2$O, yielding a low LRT statistic even when the CH$_4$ posterior itself is constrained (Figure~\ref{fig:amplify_post}). However, an alternative LRT formulation using a featureless (flat-line) null hypothesis confirms that the CH$_4$ spectral signature is intrinsically detectable by BSSL, exceeding $3\sigma$ at $\geq\times$5 amplification, consistent with the spectral amplification results shown in Figure~\ref{fig:amplify_spec}. SiO, whose spectral contribution is negligible in all scenarios, is likewise omitted from the table.}

The goal of this sensitivity analysis was to isolate the impact of instrumental capabilities (resolution, SNR) from atmospheric factors such as inherently low abundances and optically thick cloud cover. By establishing clear thresholds for the detectability and retrievability of minor atmospheric constituents at varying levels of observational sensitivity, our results provide valuable guidelines for mission planners.

\section{Discussion}
\label{sec:discussion}

JWST’s observations provide a critical empirical foundation for simulating expected outcomes with future telescopes, particularly for targets already characterised by JWST. For targets yet to be observed, empirical relationships derived from JWST data -- such as those relating to atmospheric mean molecular weight or number of scale heights (see Table~\ref{tab:transit_predictions}) -- offer valuable guidance for initial observational planning. These preliminary simulations are essential, as they enable mission teams to refine their observational strategies and allocate resources efficiently to best achieve their scientific goals.
In this context, our study evaluating Twinkle’s capabilities for atmospheric characterisation offers important initial insights into the mission’s anticipated performance in exoplanet surveys, leveraging the most recent discoveries from JWST. Our results demonstrate that, through the use of low-resolution spectroscopy and strategies involving the stacking of multiple observations, Twinkle's observations will effectively allow to determine planetary parameters and key chemical components for hundreds of exoplanets.

Focusing on transit spectroscopy of H$_2$/He-dominated atmospheres, we found that when a median SNR of 3 is achieved, key planetary parameters such as radius, atmospheric temperature and cloud heights can be constrained within meaningful ranges (e.g., GJ\,3470\,b shown in Figure \ref{fig:gj3470b_free_retrieval_results}). For chemical  species identifiable in the spectra, retrievability improves significantly as SNR increases. However, higher SNR does not always enhance the detectability of spectral features from molecules that are intrinsically weak or masked by clouds or other more dominant chemical species, emphasising the intrinsic limitations imposed by planetary atmospheres  themselves or technique adopted or Twinkle's spectral coverage and/or resolving power. In eclipse spectroscopy, Twinkle also exhibits promising capabilities, as demonstrated in our simulations for the super-Earth 55\,Cnc\,e (more details in Appendix \ref{sec:55cnce}). While characterising detailed chemical compositions remains challenging, our simulations confirm Twinkle’s ability to extract fundamental planetary parameters, such as the atmospheric thermal structure, critical for investigating internal/surface properties  and heat transport processes.
Moreover, with thousands of hours of observing time available to Twinkle's science team, continuous phase-curve observations are feasible, particularly for short-period exoplanets such as 55\,Cnc\,e. Such phase-curve data will offer insights into the three-dimensional structure of planetary atmospheres, complementing other observational approaches  \citep{2007Natur.447..183K, 2014Sci...346..838S, 2016Natur.532..207D, 2025AJ....169...32D}. Additionally, Twinkle’s ability to observe multiple transits and its high-precision timing capabilities will enhance transit ephemeris accuracy and establish more robust baselines for transit timing variation (TTV) analyses. Twinkle’s participation in TTV and transit duration variation (TDV) studies will further refine orbital parameters, improve planet mass determinations in multi-planet systems, and aid in detecting potential new exoplanets.

\subsection{H$_2$/He dominated exoplanets}
\label{sec:discuss_gasgiants}

Simulations presented for gas giants and Neptunes have significantly benefited from recent JWST observations, allowing updated assessments of their observability with Twinkle through improved understanding of their atmospheric compositions. For bright hot Jupiters, such as HD\,209458\,b, Twinkle can achieve an SNR of 10 by stacking just 5 transits at native resolution. As illustrated in Figure \ref{fig:hd209_free_post}, the improvement in retrievability for planets with prominent spectral features becomes less significant at higher SNRs. This diminished improvement for HD\,209458\,b arises because the error associated with a single transit observation by Twinkle is already relatively small. Consequently, while  further enhancements through additional observations yield limited benefits, bright targets offer an excellent opportunity to study stellar and/or atmospheric variability. For bright H$_2$/He-dominant exoplanets without prior data, an efficient strategy would be to initially perform a single transit and/or eclipse observation, conduct preliminary analysis, and then decide whether further observations are required based on initial findings or existing data from other instruments.

In contrast, for  warm H$_2$/He-dominant exoplanets orbiting fainter stars, such as WASP-107\,b, lower intrinsic observability leads to larger errors in simulated observations. Thus, increased stacking of transits and corresponding SNR improvements markedly enhance retrievability. Retrieval results for WASP-107\,b demonstrate that weaker species like Na become detectable only at higher SNR (e.g., SNR = 10). For this type of targets, an initial estimate of the required number of transits based on prior knowledge and scientific objectives would be advisable. If there is no previous knowledge,  a good approach could be to start with a limited number of transits and  then performing an initial analysis on the data acquired to design the follow-up observational plan.

Neptunes and sub-Neptune-sized planets are often cloudy and/or their atmospheres are heavier than hydrogen.
For these reasons, typically ten to twenty transits are required for adequate SNR, the lower intrinsic signal results in greater gains when improving from lower to moderate SNR. For instance, GJ\,3470\,b parameters were weakly constrained at SNR = 3 but became well constrained at SNR = 5 for molecules with multiple distinct spectral features. However, retrieval of species such as SO\textsubscript{2}, having limited spectral signatures, required higher SNR (SNR = 7). Increasing SNR beyond this point yielded diminishing returns. Similar observations were made for WASP-107\,b, where increasing observations from SNR = 5 (2 transits) to SNR = 7 (3 transits) provided more improvement than further increasing to SNR = 10 (6 transits). The diminishing marginal returns, as described by Equation \ref{eq:number_transits}, suggest an optimal observational strategy for smaller-radius planets  targeting  initially median SNR levels (3 or 5). Comparing retrieval improvements at these levels and balancing against the observational investment may help optimise follow-up observations.

In summary, Twinkle's capability to characterise exoplanet atmospheres depends primarily on stellar brightness in optical and infrared wavelengths, 
planetary and stellar radii, main atmospheric component and cloud coverage. While stellar brightness and planetary dimensions are usually well-characterised by previous observations,   cloud properties and, in the case of small planets, atmospheric mean molecular weight  may require dedicated characterisation efforts. Cloud contributions  may significantly impact parameters' retrieval, especially for minor species. \rev{For instance, in our simulated retrievals of WASP-107\,b, SO\textsubscript{2} spectral features were fully obscured by the assumed cloud opacity (see Figure~\ref{fig:wasp107b_jwst_vs_simulated}), and degeneracies between cloud-top height and planetary radius were observed.} Incorporating detailed cloud microphysical and radiative models into future simulations might allow more realistic analyses.

\subsection{Planets with Non H$_2$/He Dominated Atmosphere}
\label{sec:discuss_terr}

Atmospheric characterisation of super-Earths presents significant challenges due to their relatively small radii and heavier atmospheres, resulting in weaker observable signals. Eclipse  simulations for 55\,Cnc\,e suggest that the primary information expected from such planets is their temperature structure and constraints in atmospheric composition, distinguishable through comparisons with blackbody emission curves. If an atmosphere exists and is dominated by detectable molecules such as CO\textsubscript{2}, our simulations indicate that Twinkle can well constrain those. For bright targets such as 55\,Cnc\,e, increasing the number of eclipses beyond an initial set (10 eclipses for 55\,Cnc\,e) shows diminishing returns in parameters' retrievability.  An alternative approach for gaining deeper insights into super-Earths is phase-curve spectroscopy. The short orbital period of 55\,Cnc\,e makes continuous phase-curve monitoring feasible. We are currently conducting simulations to assess Twinkle’s potential in phase-curve observations, and these findings will be detailed in future studies.

\subsection{Stellar Activity Effects, Targets Observability and Future Works}

While stacking multiple observations to achieve higher SNR per wavelength bin is an effective strategy to enhance target observability, variability between observations introduces potential challenges. Variations in planetary atmospheric conditions between different transits or eclipses can lead to differences in the collected spectra. More significantly, stellar activity may significantly contaminate atmospheric parameters such as molecular abundances and temperature retrievals. A population study of 20 exoplanets conducted with HST by \citet{2025ApJS..276...70S} demonstrated that stellar contamination could alter molecular abundance estimates by up to six orders of magnitude and atmospheric temperatures by up to 145\%. Stellar contamination predominantly affects the optical wavelengths, which are partially covered by Twinkle. Hence, the potential influence of stellar activity must be carefully considered, especially when multiple observations are stacked. Investigating stellar activity effects on exoplanet atmospheres is one of Twinkle's core scientific themes, and a future study will explore Twinkle's capability to identify and quantify these stellar variations. \rev{In addition to astrophysical variability, instrumental systematics such as detector drift, thermal variations, and pointing jitter may affect the consistency of stacked observations. A detailed assessment of these instrumental effects will be presented in a companion study focusing on time-domain simulations.}

Another significant constraint is observational availability, influenced by Earth obstruction due to Twinkle’s orbital configuration, planetary orbital periods, and transit durations. To address this, the Twinkle team has developed an orbital planning tool, allowing members to identify available observational windows over a year. Integrating this availability with Twinkle’s simulated performance enables the development of optimised, well-informed observation strategies for target exoplanets.

However, one remaining problem is that Earth obstruction also causing Twinkle to sometimes capture partial transit or eclipse events, as described in Section \ref{sec:radiometric_snr}. Consequently, a more realistic estimate of the number of required observations should incorporate an observing efficiency factor, as demonstrated by \citet{2024MNRAS.530.2166B}. Meanwhile, the Twinkle team is actively developing methodologies for effectively stacking multiple partial transits or eclipses to maximise SNR. Comparable techniques have already been demonstrated for combining multiple nights of data with the same instrumental setup in both transmission and emission studies from ground-based instruments \citep[e.g.,][]{2021Natur.592..205G, 2022AJ....163..107K, 2023A&A...674A..58S}. A simulation by \citet{2024A&A...683A.244B} specifically addresses the compromises required when maximising SNR through stacked observations for transiting exoplanets.

Several studies underline the importance of combining multiple observations for the future characterisation of Earth-like and super-Earth exoplanets, even for powerful observatories such as JWST \citep{2019A&A...624A..49W}. Indeed, recent analysis by \citet{2023NatAs...7.1317L} involved combining two JWST transit observations for improved results. Historically, enhancing SNR through stacking multiple observations has often been impractical due to limited telescope availability. However, as a commercially driven, rapidly deployable mission, Twinkle can offer substantial, on-demand telescope time, potentially thousands of hours for each scientific theme. Therefore, stacking multiple observations as availability permits is a viable and efficient strategy for the Twinkle science community, provided that stacking does not introduce significant data distortions.

\section{Conclusions}

Here we present updates on the Twinkle space missions simulated performance for 
characterising exoplanetary atmospheres,
given the latest design choices of the spacecraft and payload and new insight into the atmospheric composition and structure of select exoplanets provided by JWST’s observations.
In addition, we have updated the list of exoplanet candidates suitable for characterisation based on currently confirmed planets and TOIs, and incorporating updated planetary and stellar parameters as available in the literature. 

More specifically, leveraging recent discoveries from JWST, we have provided more realistic spectral simulations across various exoplanet types and assessed the retrievability of their atmospheric parameters and chemical abundances at different achievable SNR levels. 
Retrieval  analyses for HD\,209458\,b, WASP-107\,b, and GJ\,3470\,b have indicated how increased observational investment to reach  higher  SNR improves the retrievability of the atmospheric parameters and chemical species for different planetary categories. At Twinkle’s native resolution, already at low median SNR values of 3 or 5, key atmospheric properties, such as temperature  and cloud altitudes, are well constrained, alongside  major molecular species. Increasing the median SNR to 7 or 10 further improves constraints on minor species, exemplified by the successful retrieval of SO\textsubscript{2} in GJ\,3470\,b. For small planets with atmospheres heavier than hydrogen, such as 55\,Cnc\,e, emission retrieval studies revealed that after stacking ten eclipses, temperature structure and dominant atmospheric constituents could be  constrained. Additional eclipse observations  showed limited improvement, however 55 Cnc e would be an excellent candidate for phase-curve observations with Twinkle. This example showcases that tailored observational strategies are necessary for different exoplanet types  to ensure an efficient use of the telescope time.

Our analysis indicates that approximately 100 exoplanets are suitable for in depth atmospheric studies and over 300 exoplanets are good candidates for characterisation with Twinkle.

Future studies will build upon this work to better inform the observational planning of the Twinkle's exoplanet survey. In particular we plan to include more realistic assumptions  of target observability due to Earth obstruction and quantify the impact of stellar activity in our simulations.
We also plan to refine our simulations by integrating more advanced cloud formation models and to assess the detectability of other atmospheric features. Additionally, the radiometric model will be continuously updated alongside instrumental or operational developments to ensure precise noise estimates. We will also broaden the scope of simulations to include phase-curve observations, exploring expected outcomes for other scientific themes and their mutual refinements.

In the next five years, numerous space- and ground-based observatories dedicated to exoplanetary science research will become operational. BSSL-Twinkle uniquely complements these efforts by offering the science team full flexibility in the selection of the list of targets and observational strategy. To support informed decision-making by the Twinkle science team, BSSL is committed to clearly showcasing the telescope’s anticipated scientific outcomes. 

\section*{Acknowledgements}
We gratefully acknowledge insightful discussions and inputs from members of the Twinkle Science Team, and their participation in Twinkle Science Conference. \rev{This work made use of data and results from observations with the NASA/ESA/CSA James Webb Space Telescope, as reported in the published literature. The JWST data were obtained from the Mikulski Archive for Space Telescopes at the Space Telescope Science Institute, which is operated by the Association of Universities for Research in Astronomy, Inc., under NASA contract NAS 5-03127 for JWST.}

\section*{Data Availability}
The data underlying this article will be shared on reasonable request to the corresponding author.

\section*{Conflict of Interest}
\rev{The authors declare no conflict of interest.}



\bibliographystyle{rasti}
\bibliography{rasti} 




\appendix
\section{Noise Contributions} \label{appendix:noise}
\rev{The noise component plots presented in this section illustrate how the different noise sources considered by the Radiometric Tool vary across each target. For all targets simulated in this work, the noise budget remains in the photon-noise-dominated regime, with all noise contributions increasing towards the edges of each spectral channel. For relatively bright targets such as HD\,209458\,b and 55\,Cnc\,e, read noise begins to exceed photon noise at the longer wavelengths covered by channel~1, as the detector requires more reads to avoid saturation.}

\rev{The radiometric simulations presented in this study adopt the instrument parameters
summarised in Table~\ref{tab:instrument_params}. In its current design, the telescope has an effective
collecting area of $A_\mathrm{tel} = 0.146$\,m$^2$. The optical train consists of
four mirror surfaces and a dichroic beam splitter in the common path, followed by
a dispersing prism and three fold mirrors in each channel. All mirror surfaces are
assumed to have a reflectivity of 98\%, while the dichroic and prism each transmit
$\sim$90\% and $\sim$80\%, respectively.
The detector quantum efficiency is wavelength-dependent, ranging from
75\% to 95\% across the 0.5--4.5\,$\mu$m bandpass.}

\begin{figure*}
    \centering
    \begin{subfigure}{0.8\textwidth}
        \centering
        \includegraphics[width=\textwidth]{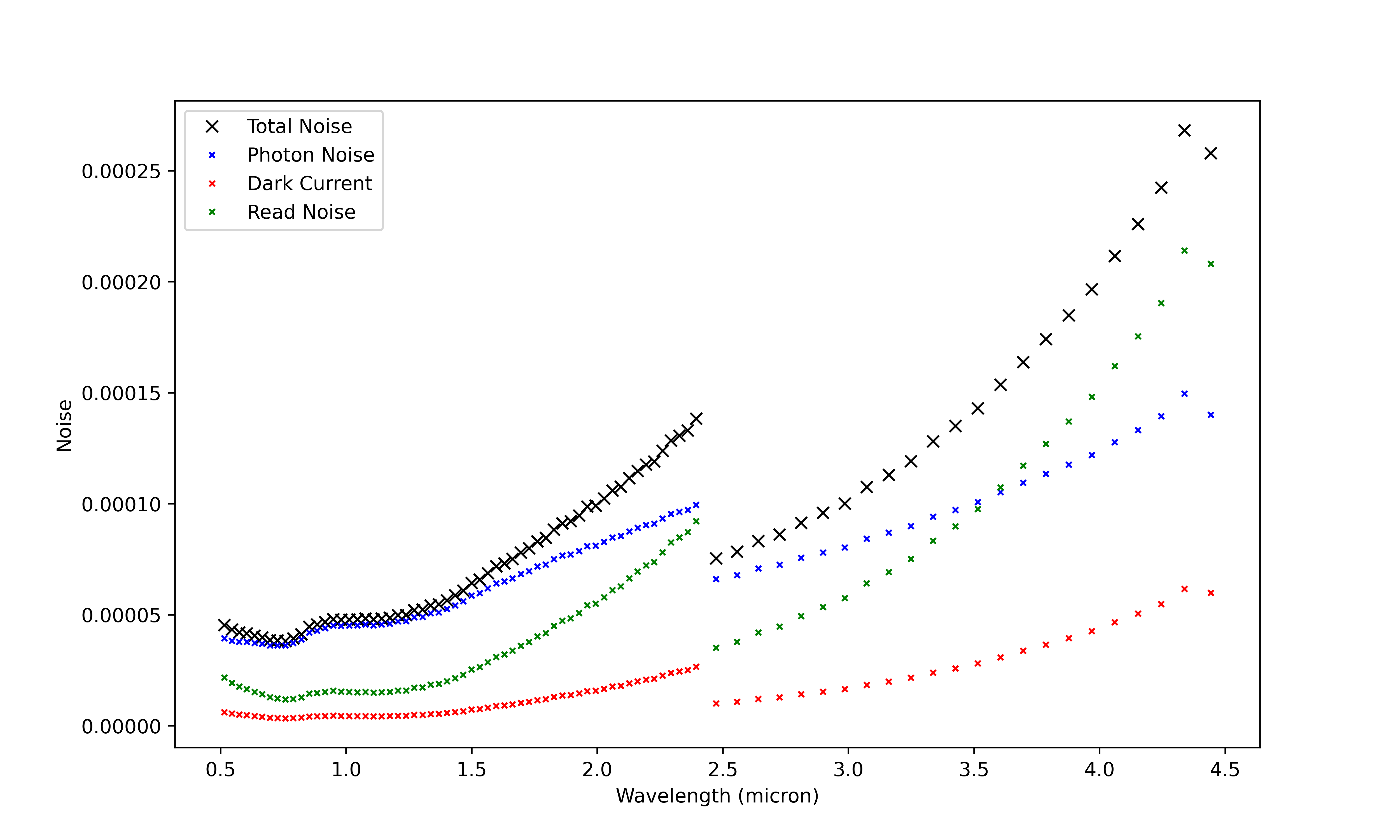}
        \caption{HD\,209458\,b}
        \label{fig:hd209_noise}
    \end{subfigure}
    \vspace{0.5cm}
    \begin{subfigure}{0.8\textwidth}
        \centering
        \includegraphics[width=\textwidth]{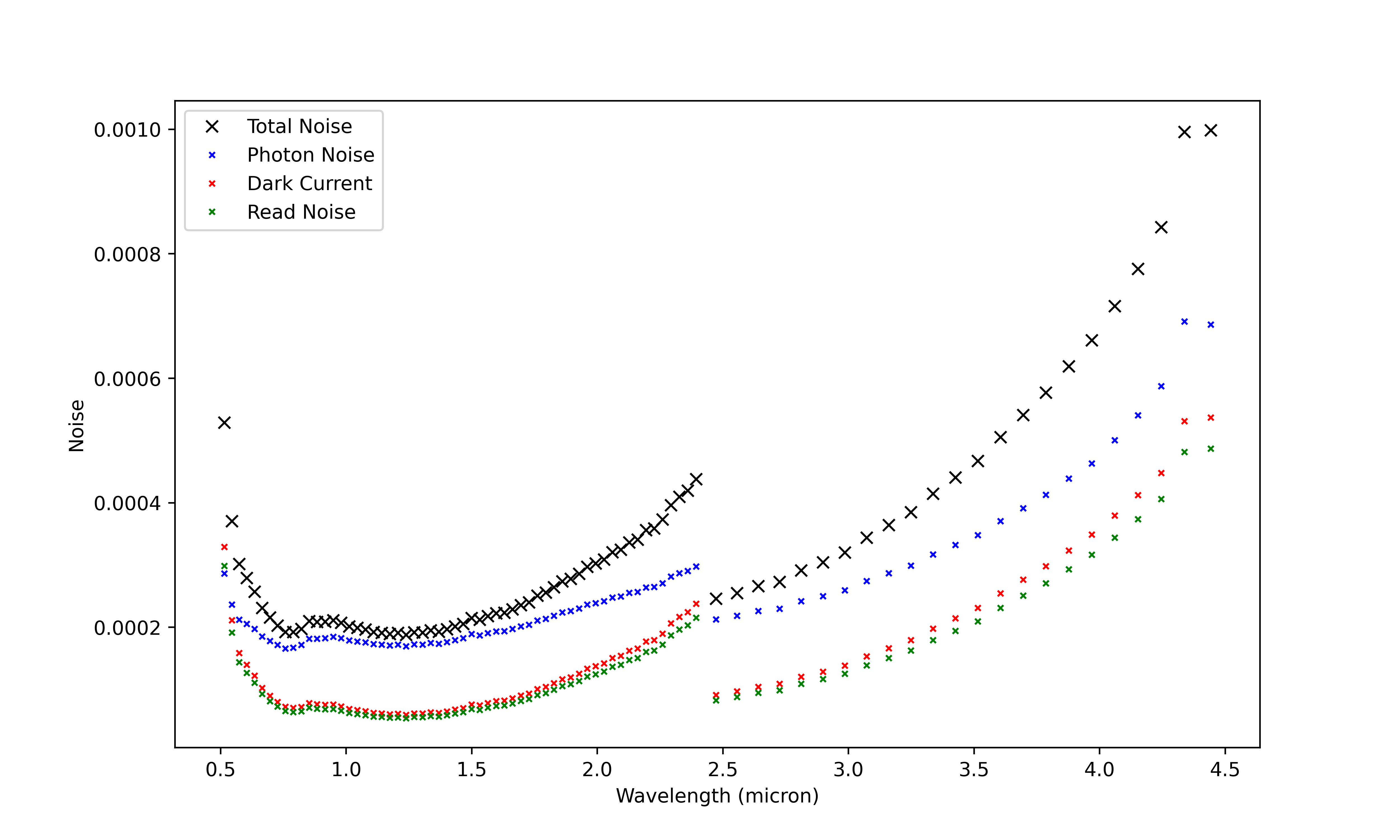}
        \caption{WASP-107\,b}
        \label{fig:wasp107b_noise}
    \end{subfigure}
    \caption{\rev{Noise component breakdown for a single transit observation at Twinkle's native spectral resolution. The total noise (black) is decomposed into photon noise (blue), dark current (red), and read noise (green). The discontinuity at 2.43\,$\mu$m marks the boundary between channel~0 and channel~1. (a)~For the bright host star of HD\,209458\,b ($K_\mathrm{mag} \approx 6.3$), read noise becomes the dominant contributor in channel~1 because the detector requires more frequent reads to avoid well saturation. (b)~For the fainter WASP-107\,b ($K_\mathrm{mag} \approx 9.0$), longer integration times are feasible and photon noise dominates across both channels; the overall noise level is correspondingly higher.}}
    \label{fig:noise_hd209_wasp107}
\end{figure*}

\begin{figure*}
    \centering
    \begin{subfigure}{0.8\textwidth}
        \centering
        \includegraphics[width=\textwidth]{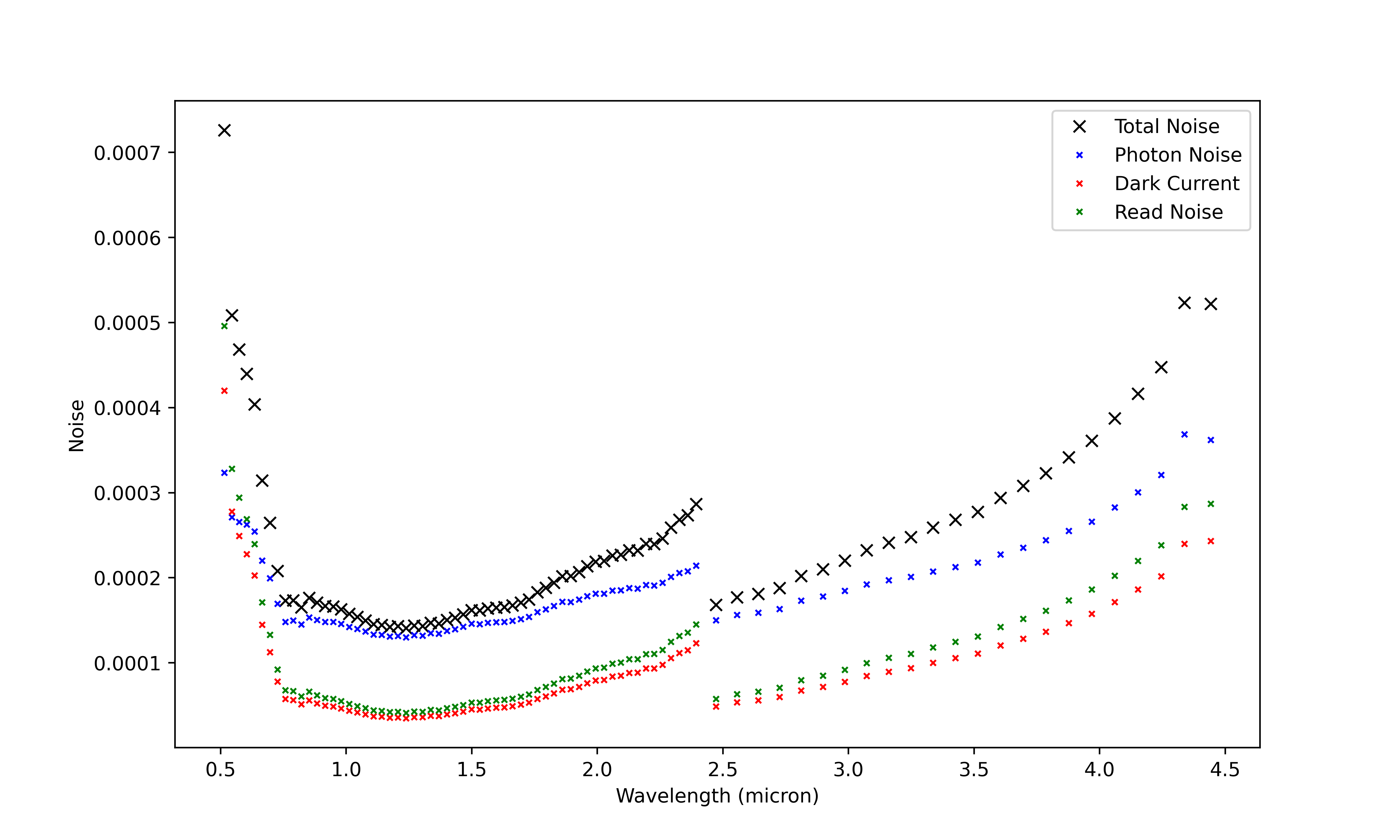}
        \caption{GJ\,3470\,b}
        \label{fig:gj3470_noise}
    \end{subfigure}
    \vspace{0.5cm}
    \begin{subfigure}{0.8\textwidth}
        \centering
        \includegraphics[width=\textwidth]{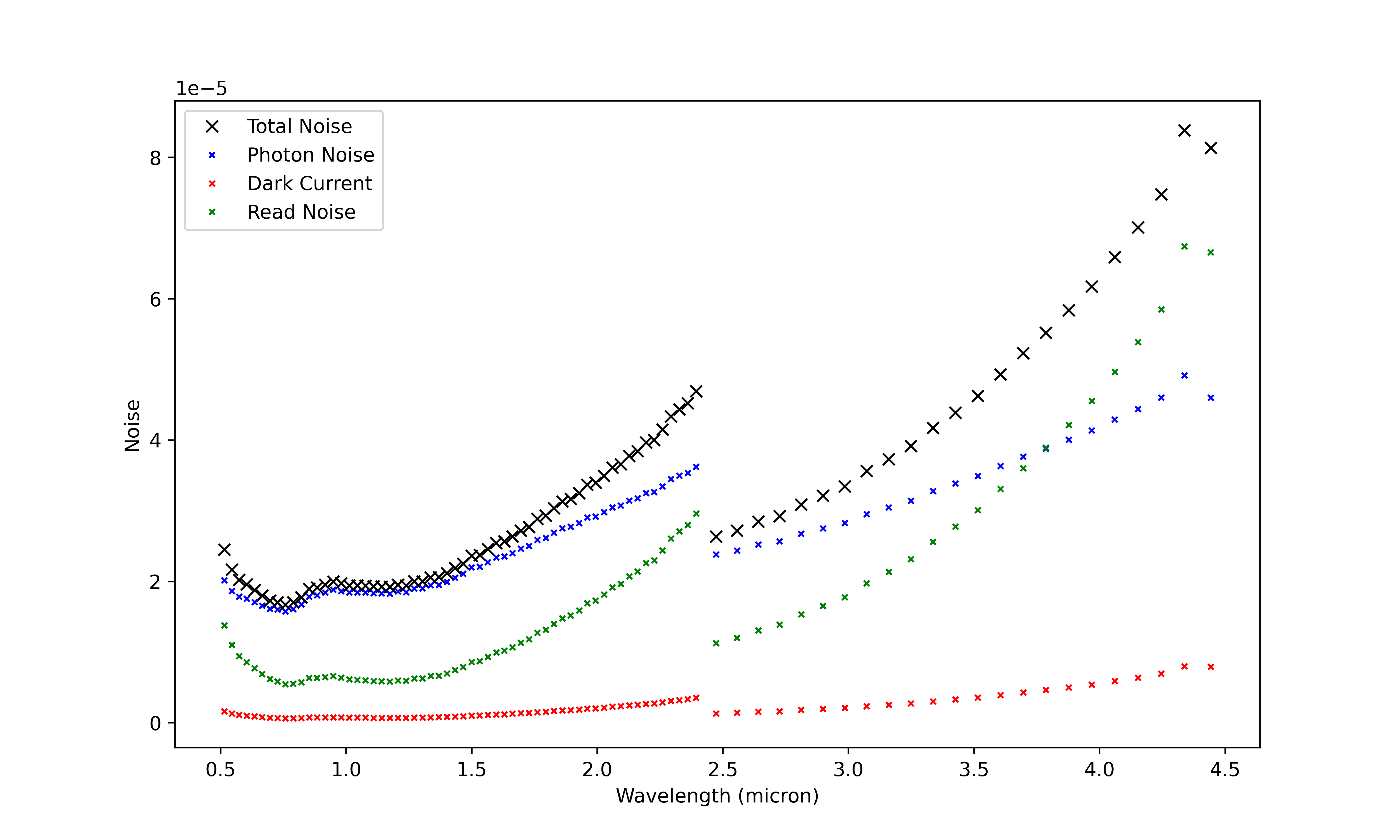}
        \caption{55\,Cnc\,e}
        \label{fig:55cnce_noise}
    \end{subfigure}
    \caption{\rev{Same as Figure~\ref{fig:noise_hd209_wasp107} but for GJ\,3470\,b and 55\,Cnc\,e. (a)~For GJ\,3470\,b ($K_\mathrm{mag} \approx 7.99$), photon noise dominates across both channels, with a noise level intermediate between the bright and faint targets. (b)~55\,Cnc\,e is the brightest target in our sample ($K_\mathrm{mag} \approx 4.0$); the detector saturates rapidly, requiring very short integration times and a large number of reads per exposure. Consequently, read noise dominates much of channel~1, more pronounced than for HD\,209458\,b, while the absolute noise level is the lowest among all four targets.}}
    \label{fig:noise_gj3470_55cnce}
\end{figure*}

\rev{The point spread function (PSF) in channel~0 is treated as diffraction-limited,
while channel~1 assumes a wavefront error of $\mathrm{WFE}_\mathrm{rms} = 0.28$\,$\mu$m.
Pixel-based PSF models are used for both channels. The detectors operate at 90\,K
with a read noise of 25 counts per read, a dark current of 35 counts\,s$^{-1}$,
and a well depth of 72{,}000 counts (80\% of which is usable to minimise
non-linear detector effects). For bright targets such as HD\,209458\,b and
55\,Cnc\,e, shorter integration times are required to avoid well saturation,
leading to a higher number of detector reads per exposure and consequently
elevated read noise contributions (Figures~\ref{fig:hd209_noise} and
\ref{fig:55cnce_noise}). For fainter targets (WASP-107\,b, GJ\,3470\,b),
longer integrations are feasible, and photon noise dominates across the
full spectral range.}

\rev{Line-of-sight pointing jitter is included in the noise model using
a pre-computed jitter profile derived from ground-test pointing performance data.
The out-of-transit noise overhead is accounted for by a factor
$\gamma_\mathrm{noise} = 1.67$, corresponding to an out-of-transit baseline
equal to 1.5 times the transit duration. The zodiacal background is modelled
using a position-dependent sky map with a nominal scaling factor of 2.5.
Additional systematic noise contributions (e.g.\ gain noise, detector drift,
thermal variations) are not included in the present radiometric model but will
be addressed in a forthcoming time-domain simulation study.}

\begin{table}
\centering
\caption{\rev{Summary of Twinkle instrument parameters assumed in the radiometric simulations.}}
\label{tab:instrument_params}
\rev{
\begin{tabular}{lc}
\hline
\textbf{Parameter} & \textbf{Value} \\
\hline
Collecting area ($A_\mathrm{tel}$) & 0.146\,m$^2$ \\
Channel 0 range & 0.5--2.43\,$\mu$m ($R \leq 70$) \\
Channel 1 range & 2.43--4.5\,$\mu$m ($R \leq 50$) \\
Mirror reflectivity (per surface) & 98\% \\
Dichroic transmission & 90\% \\
Prism transmission & 80\% \\
Detector QE & 75--95\% \\
Read noise & 25 counts \\
Dark current & 35 counts\,s$^{-1}$ \\
Well depth & 72{,}000 counts \\
Detector temperature & 90\,K \\
Optics temperature & 100--160\,K \\
PSF (Ch 0) & Diffraction-limited \\
PSF (Ch 1) & $\mathrm{WFE}_\mathrm{rms} = 0.28$\,$\mu$m \\
\hline
\end{tabular}
}
\end{table}

\rev{The observation stacking equation (Equation~\ref{eq:number_transits}) assumes
that the noise contributions from successive transits or eclipses are uncorrelated.
While this assumption is appropriate for photon noise and random detector noise,
systematic effects such as instrumental drift, thermal cycling, or stellar variability
may introduce correlated noise components that do not average down as $\sqrt{N}$.
Experience with \textit{HST}/WFC3 transit spectroscopy suggests that systematic
noise floors of $\sim$20--30\,ppm can be reached for bright targets after
careful detrending \citep{2013ApJ...774...95D}. A detailed assessment of
Twinkle's systematic noise floor, incorporating time-domain instrumental
effects, will be presented in a companion study.}

\section{Retrieval Posterior Plots} \label{appendix:posteriors}

The posterior plots presented in this section of Appendix illustrate the statistical distributions of retrieved atmospheric parameters obtained from our retrieval analyses. Each posterior plot summarises the likelihood distributions of key atmospheric properties, including planetary radius, equilibrium temperature, cloud-top pressure, and molecular abundances, under various observational conditions (e.g., different SNR levels). The contours indicate parameter uncertainties and correlations. Narrower, well-defined distributions imply robust retrievals and tightly constrained parameters, whereas broader, elongated distributions reflect higher uncertainties or parameter degeneracies. These visualisations highlight how observational strategies, such as varying SNRs or the number of stacked transits, impact the precision and reliability of atmospheric characterisation.

\begin{figure*}
    \centering
    \includegraphics[width=0.7\textwidth]{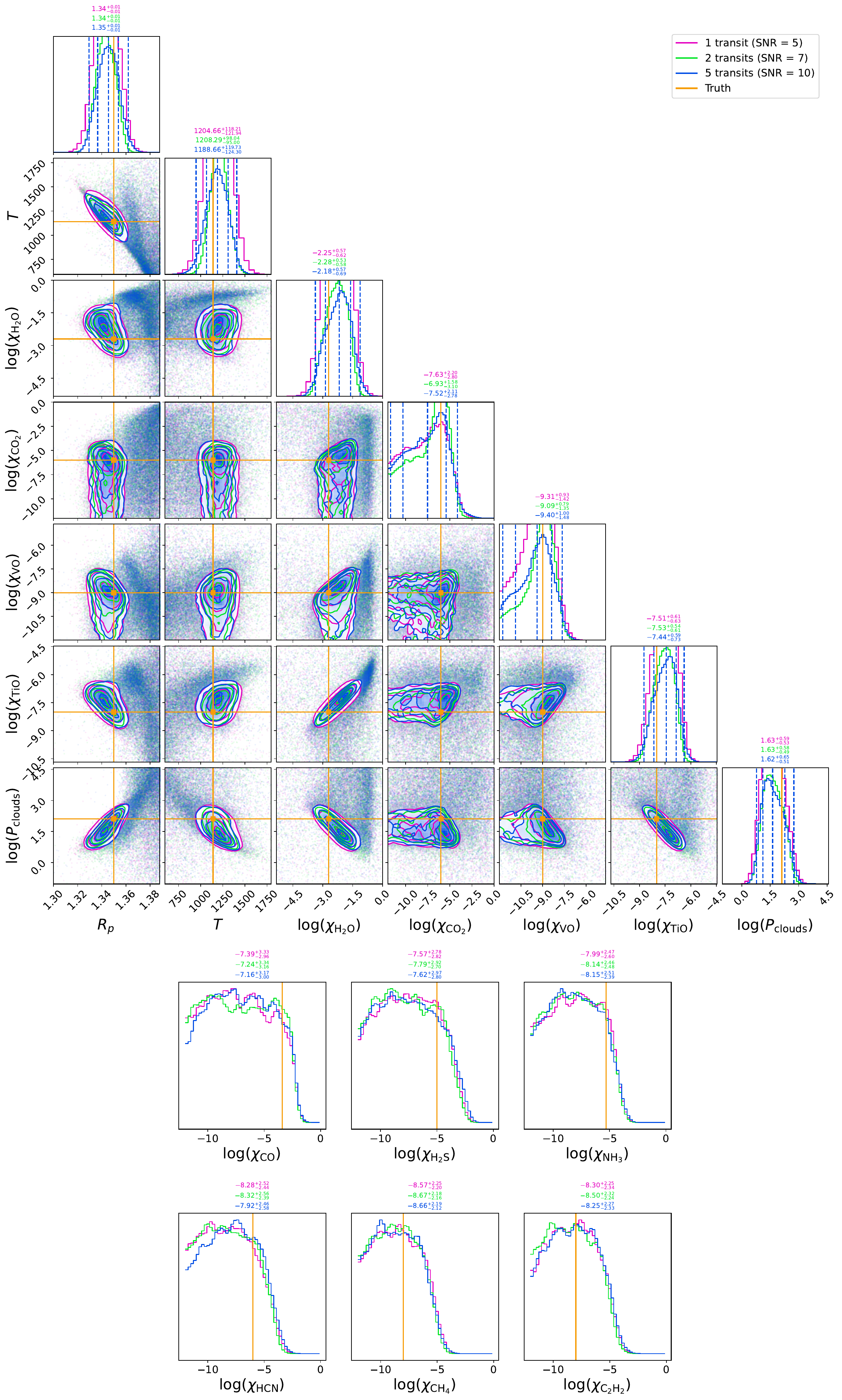} 
    \caption{Posterior distributions of retrieved parameters from free retrieval analysis of HD\,209458\,b. Planet parameters and molecules visible in the spectrum are shown in the corner plot. Invisible molecules with retrieved upper limits are shown in the histograms at the bottom. The numbers above each block represent the median value of the posteriors.}
    \label{fig:hd209_free_post}
\end{figure*}

\begin{figure*}
    \centering
    \includegraphics[width=0.7\textwidth]{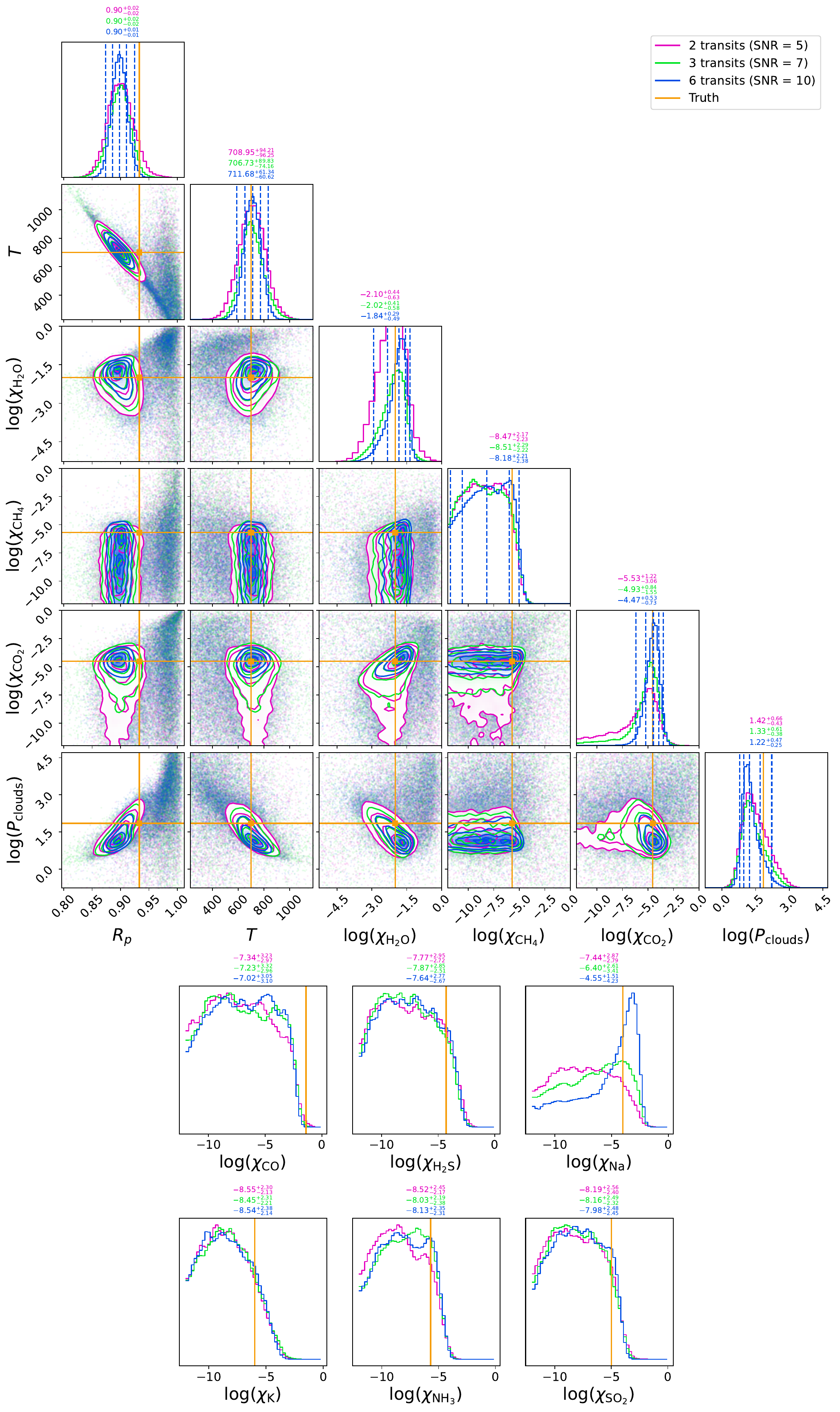}
    \caption{Posterior distributions of retrieved parameters from free retrieval analysis of WASP 107 b. Planet parameters and molecules visible in the spectrum are shown in the corner plot. Invisible molecules with retrieved upper limits are shown in the histograms at the bottom.}
    \label{fig:wasp107b_free_retrieval_results}
\end{figure*}

\begin{figure*}
    \centering
    \includegraphics[width=0.7\textwidth]{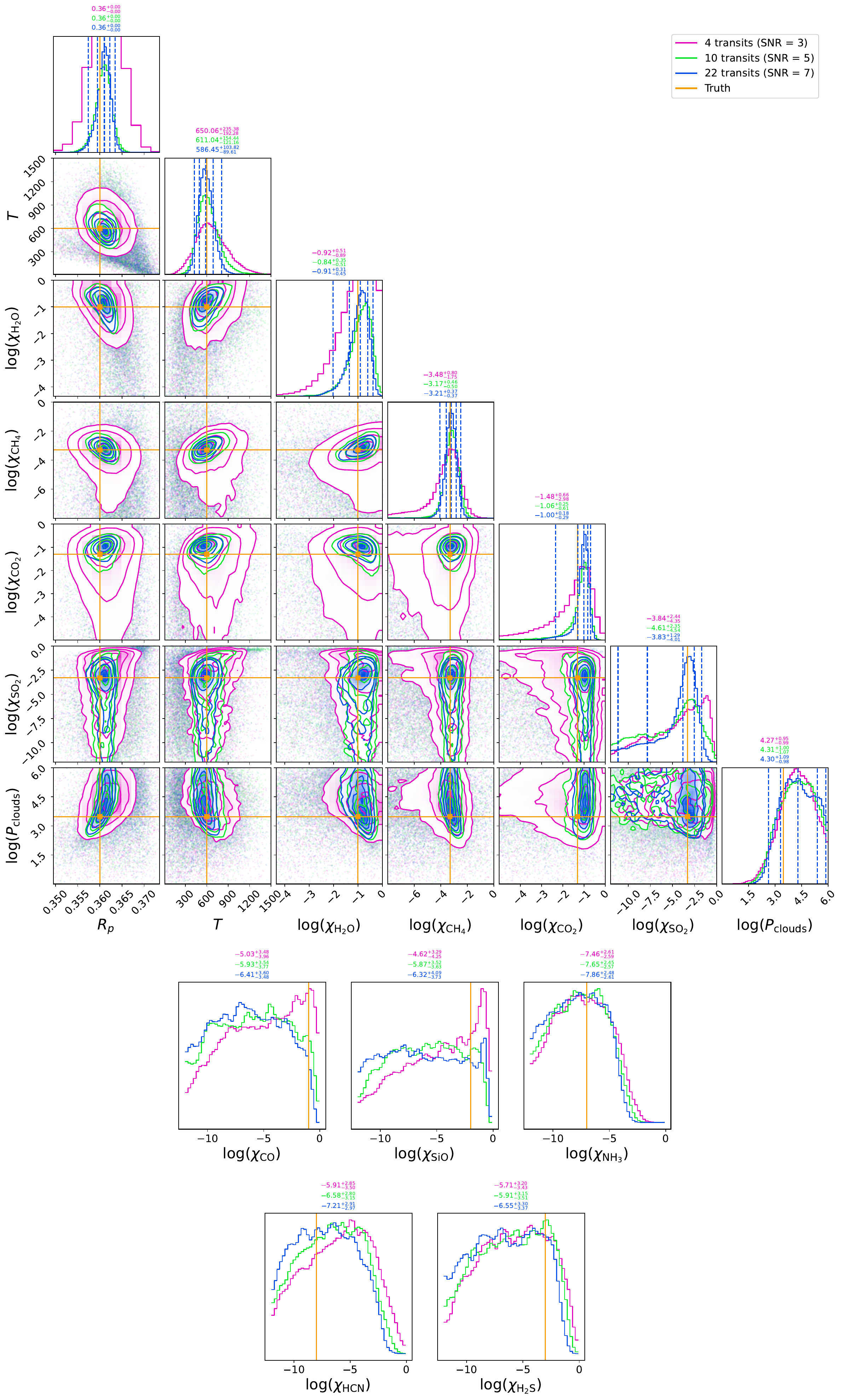}
    \caption{Posterior distributions of retrieved parameters from free retrieval analysis of GJ\,3470\,b. Planet parameters and molecules visible in the spectrum are shown in the corner plot. Invisible molecules with retrieved upper limits are shown in the histograms at the bottom.}
    \label{fig:gj3470b_free_retrieval_results}
\end{figure*}

\begin{figure*}
    \centering
    \includegraphics[width=0.7\textwidth]{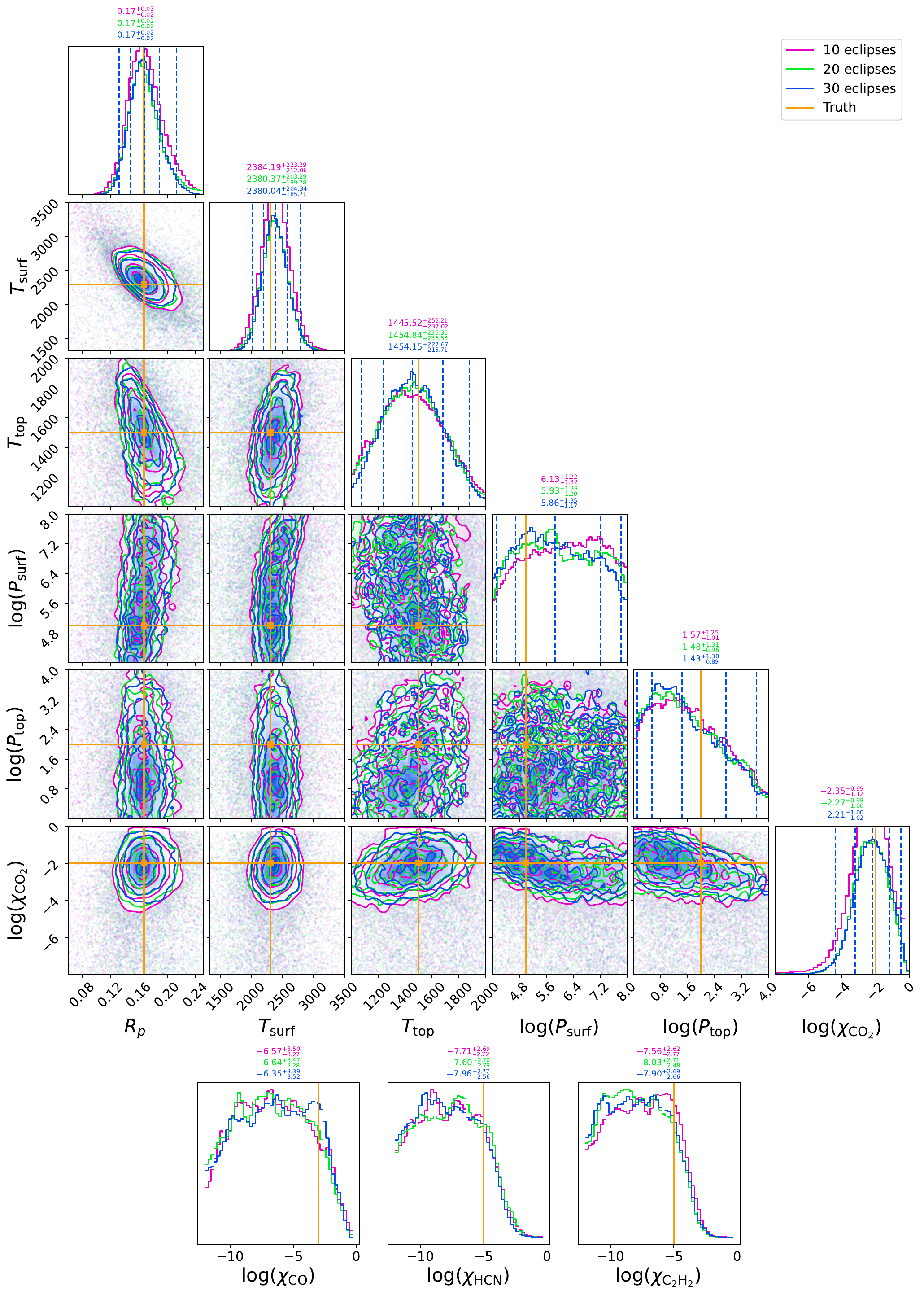}
    \caption{Posterior distributions of retrieved parameters from free retrieval analysis of 55\,Cnc\,e. Planet parameters and molecules visible in the spectrum are shown in the corner plot. Invisible molecules with retrieved upper limits are shown in the histograms at the bottom.}
    \label{fig:55cnce_retri}
\end{figure*}

\clearpage
\section{Mass Estimates for TESS Object of Interests}
\label{sec:A_mass}
\begin{figure*}
    \centering
    \includegraphics[width=0.7\textwidth]{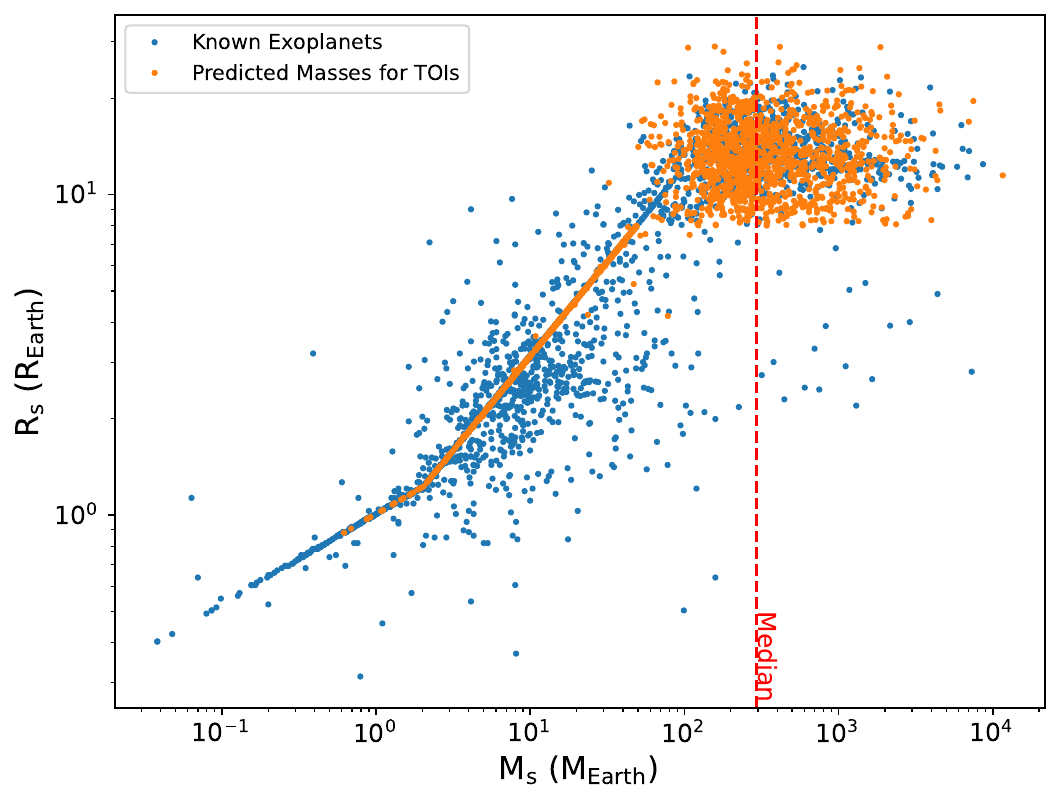}
    \caption{Mass-Radius relationship comparison between all transiting known exoplanets and TOIs in Twinkle's FoR with mass festimate based on radius. The red dashed vertical line in the plot represents the median mass of known exoplanets with radii larger than 8 Jupiter radii, which is counted as the mean value for the log-normal distribution of TOIs mass estimate.}
    \label{fig:mass_prediction}
\end{figure*}

In order to estimate masses for TOIs with radius available, we utilised \textit{Forecaster} tool developed by \citet{2017ApJ...834...17C}, a code that predict mass from radius for objects. We take the maximum radius of known exoplanets as an edge, and treat those with larger radii as dwarfs and ignore them in this study. Mass estimate using \textit{Forecaster} tool for TOIs with radius less than 8 Earth radii is being adopted as they align with the known exoplanets distribution as shown in Figure \ref{fig:mass_prediction}. For TOIs with higher radius, we assume an asymmetric log-normal distribution to randomly set the mass, with the mean value set as the median value of known exoplanets with radius larger than 8 Earth radii. Resultant mass distribution for TOIs in Twinkle's FoR is shown in Figure \ref{fig:mass_prediction}. 

\section{Likelihood Ratio Test for Molecular Detectability}
\label{appendix:lrt}

\rev{To quantify the statistical significance of molecular detections, we employ the Likelihood Ratio Test (LRT) following the methodology described by \citet{2013MNRAS.436.2974T}. For each molecule under consideration, the alternative hypothesis $H_1$ corresponds to the full forward model containing all retrieved species, while the null hypothesis $H_0$ corresponds to the model with the test molecule removed. For each hypothesis, Monte Carlo simulations with 100,000 iterations generate synthetic observations by adding Gaussian noise (drawn from the per-bin radiometric uncertainties) to the corresponding model spectrum. For each realisation, the likelihood ratio statistic $\Delta = -2\ln\mathcal{L}_0 + 2\ln\mathcal{L}_1$ is computed, yielding distributions of $\Delta$ (data generated under $H_0$) and $\Delta'$ (data generated under $H_1$). A molecule is classified as a ``Detection'' when median($\Delta'$) $>$ mean($\Delta$) $+ 3\sigma(\Delta)$, where $\sigma(\Delta)$ is the standard deviation of the null distribution. The detection significance, expressed in units of $\sigma$, is reported in the results tables. Molecules falling below the $3\sigma$ threshold are classified as ``Upper Limit'', indicating that their abundances are either prior-dominated or only weakly constrained by the data.}

\rev{For emission spectra (55\,Cnc\,e), we adopt the same molecule-removal framework: $H_1$ is the full emission model containing all retrieved species, while $H_0$ is the full model with the test molecule removed. This approach is more conservative than the original \citet{2013MNRAS.436.2974T} emission formulation (which uses a featureless blackbody as $H_0$), because it tests whether a molecule's spectral signature is distinguishable in the presence of all other atmospheric constituents, rather than merely deviating from a smooth continuum.}

\rev{To ensure robustness against unreliable MAP estimates for poorly constrained parameters, we perform the LRT using both the MAP and posterior median values for each molecule. A molecule is classified as a ``Detection'' only if both tests independently exceed the $3\sigma$ threshold. The more conservative (lower) of the two significance values is reported in the results tables.}

\begin{figure*}
    \centering
    \includegraphics[width=\textwidth]{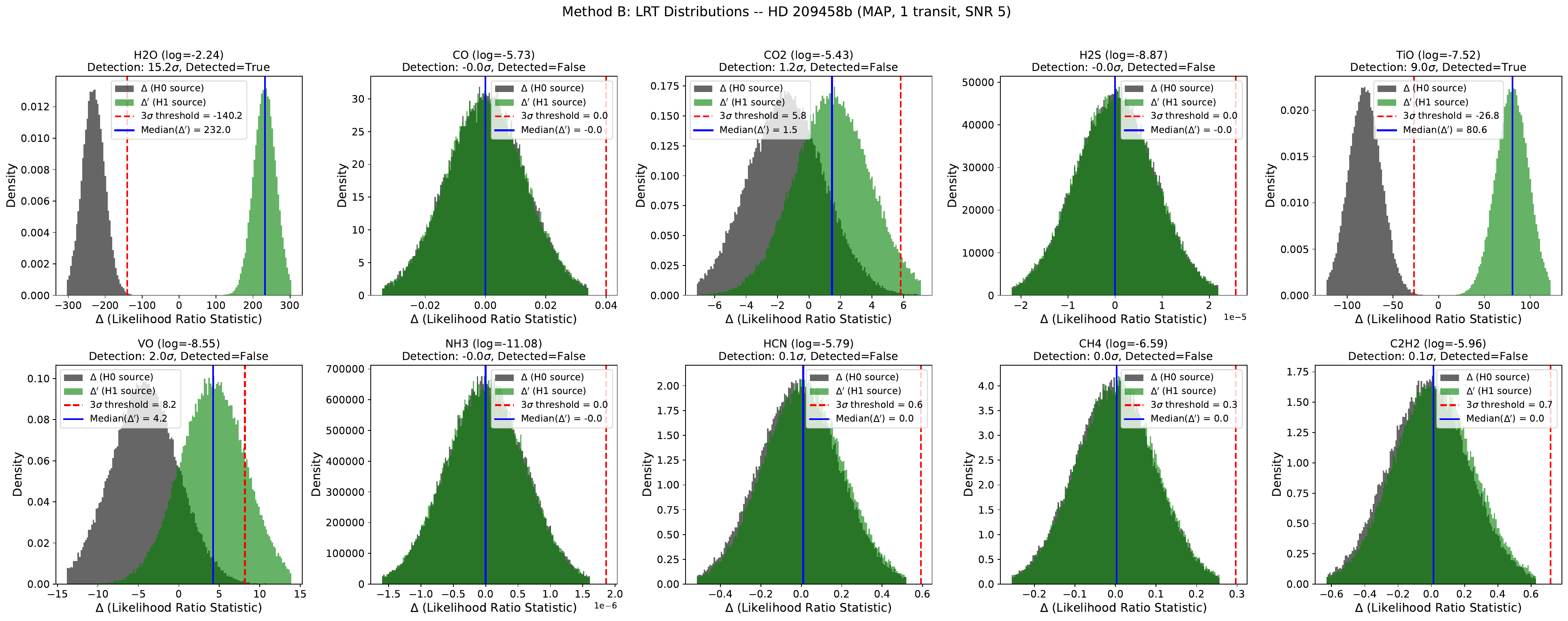}
    \caption{\rev{LRT $\Delta$ and $\Delta'$ distributions for HD\,209458\,b (1 transit, SNR\,5). Each panel shows the null distribution $\Delta$ (black) and alternative distribution $\Delta'$ (green) for a single molecule removal test. The red dashed line marks the $3\sigma$ detection threshold; the blue solid line marks the median of $\Delta'$. Detection significance in units of $\sigma$ is indicated in each panel title.}}
    \label{fig:lrt_hd209458b}
\end{figure*}

\begin{figure*}
    \centering
    \includegraphics[width=\textwidth]{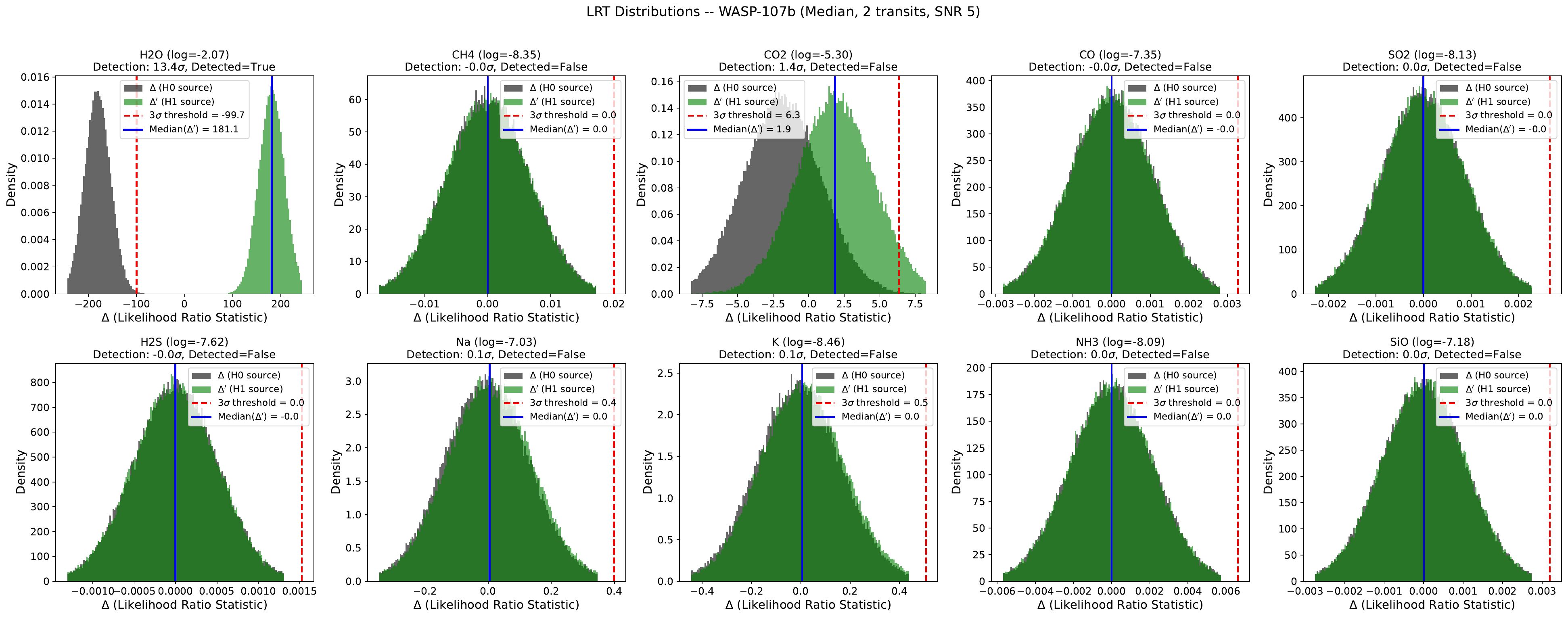}
    \caption{\rev{LRT $\Delta$ and $\Delta'$ distributions for WASP-107\,b (2 transits, SNR\,5). Panel layout and colour scheme as in Figure~\ref{fig:lrt_hd209458b}.}}
    \label{fig:lrt_wasp107b}
\end{figure*}

\begin{figure*}
    \centering
    \includegraphics[width=\textwidth]{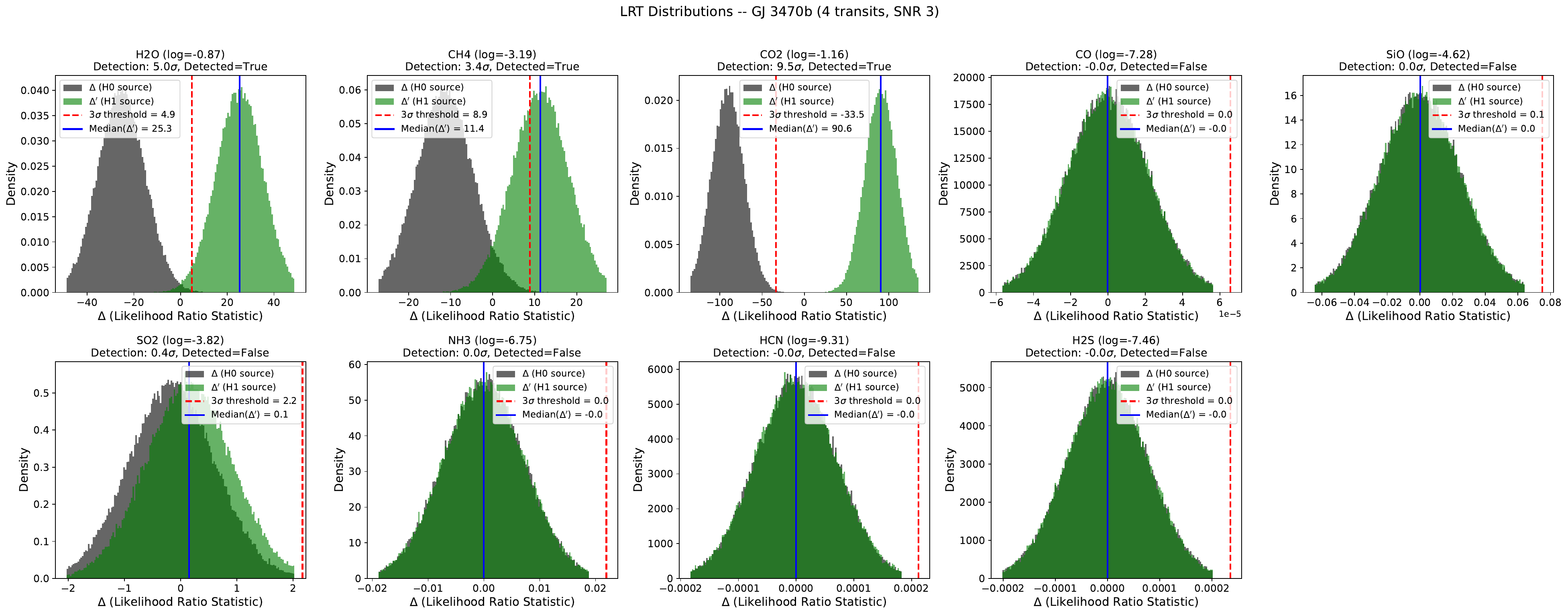}
    \caption{\rev{LRT $\Delta$ and $\Delta'$ distributions for GJ\,3470\,b (4 transits, SNR\,3). Panel layout and colour scheme as in Figure~\ref{fig:lrt_hd209458b}.}}
    \label{fig:lrt_gj3470b}
\end{figure*}

\begin{figure*}
    \centering
    \includegraphics[width=\textwidth]{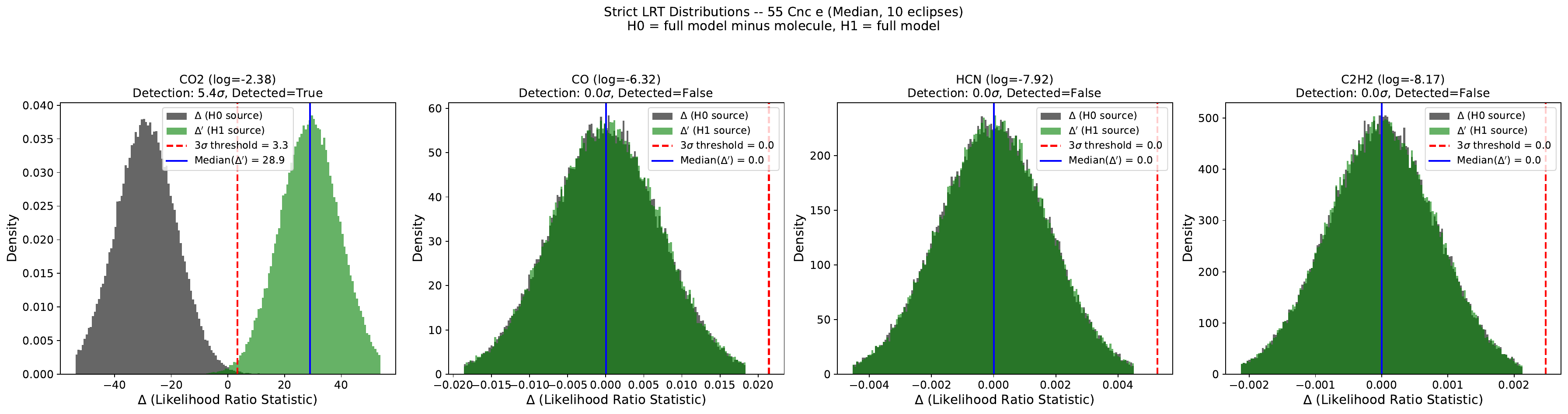}
    \caption{\rev{LRT $\Delta$ and $\Delta'$ distributions for 55\,Cnc\,e (10 eclipses). For this emission target, $H_1$ is the full emission model and $H_0$ is the full model with the test molecule removed. Panel layout and colour scheme as in Figure~\ref{fig:lrt_hd209458b}.}}
    \label{fig:lrt_55cnce}
\end{figure*}

\rev{Figures~\ref{fig:lrt_hd209458b}--\ref{fig:lrt_55cnce} show the resulting $\Delta$ (black) and $\Delta'$ (green) distributions for each molecule tested. The red dashed line indicates the $3\sigma$ detection threshold, while the blue solid line marks the median of $\Delta'$. A molecule is detected when the blue line exceeds the red line.}

\bsp	
\label{lastpage}
\end{document}